\documentclass[aps,pra,twocolumn,eqsecnum,superscriptaddress,showpacs]{revtex4-1}
\usepackage{graphicx,color}
\usepackage{amsmath}
\usepackage{amssymb}
\usepackage{bm}

\newcommand{\Tr}{\mathop{\text{Tr}}\nolimits}
\newcommand{\ket}[1]{|{#1}\rangle}
\newcommand{\bra}[1]{\langle{#1}|}

\definecolor{dgreen}{rgb}{0,0.5,0}

\definecolor{dblue}{rgb}{0,0,0.6}

\definecolor{dred}{rgb}{0.784,0,0}

\definecolor{delete}{cmyk}{0.5,0,0,0}

\newcommand{\DEL}[1]{}

\begin{document}
\title{Quantum Estimation via Sequential Measurements}
\author{Daniel Burgarth}
\affiliation{Department of Mathematics, Aberystwyth University, SY23 3BZ Aberystwyth, United Kingdom}
\author{Vittorio Giovannetti}
\affiliation{NEST, Scuola Normale Superiore and Istituto Nanoscienze-CNR,  
I-56126 Pisa, Italy}
\author{Airi N. Kato}
\affiliation{Department of Physics, University of Tokyo, Tokyo 113-0033, Japan}
\author{Kazuya Yuasa}
\affiliation{Department of Physics, Waseda University, Tokyo 169-8555, Japan}
\date{\today}
\begin{abstract}
The problem of estimating a parameter of a quantum system through a series of measurements performed sequentially on a quantum probe is analyzed in the general setting where the underlying statistics is explicitly non-i.i.d.
We present a generalization of the central limit theorem in the present context, which under fairly general assumptions shows that as the number $N$ of measurement data increases the probability distribution of functionals of the data (e.g., the average of the data) through which the target parameter is estimated becomes asymptotically normal and independent of the initial state of the probe. 
At variance with the previous studies  [M. Gu\c{t}\u{a}, Phys.\ Rev.\ A \textbf{83}, 062324 (2011); M. van Horssen and M. Gu\c{t}\u{a}, J. Math.\ Phys.\ \textbf{56}, 022109 (2015)] we take a diagrammatic approach, which allows one to compute not only the leading orders in $N$ of the moments of the average of the data but also those of the correlations among subsequent measurement outcomes.
In particular our analysis points out that the latter, which are not available in usual i.i.d.\ data, can be exploited in order to improve the accuracy of the parameter estimation.
An explicit application of our scheme is discussed by studying how the temperature of a thermal reservoir can be estimated via sequential measurements on a quantum probe in contact with the reservoir.
\end{abstract}
\maketitle

\section{Introduction}
Seeking the most efficient way to recover the value of a parameter $g$ encoded in the state $\rho_g$ of a quantum system is the fundamental problem of a branch of quantum information technologies \cite{BENSHOR}, which goes under the name of quantum metrology \cite{ref:QuantumMetrologyVittorio,ref:MetrologyNaturePhoto}.
It goes without mentioning that this topic has applications in a variety of different research areas, ranging e.g.\ from the interferometric estimation of the phase shifts induced by gravitational waves \cite{VIRGO}, high-precision quantum magnetometry \cite{AUZ,MAG1}, to remote probing of targets.

\begin{figure}[t]
\begin{tabular}{l}
(a)\\[-3.5truemm]
\multicolumn{1}{c}{\includegraphics[scale=0.52]{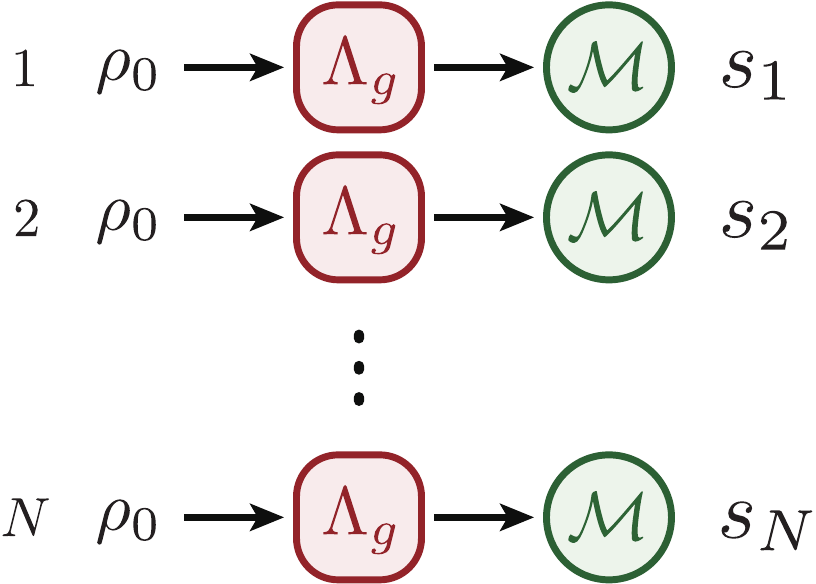}}\\[3.5truemm]
(b)\\
\includegraphics[scale=0.52]{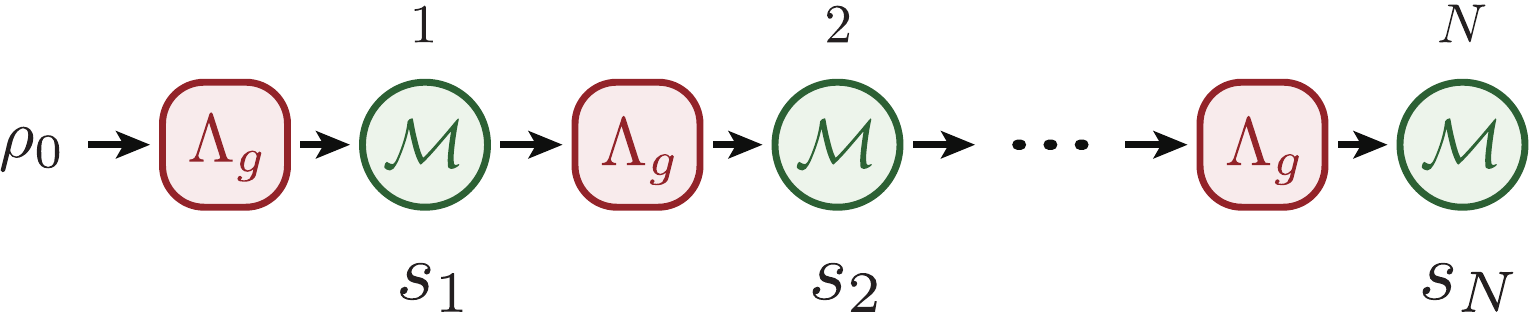}
\end{tabular}
\caption{(a) The standard strategy for estimating a parameter $g$ of a quantum system, where measurement data $\{s_1,\ldots,s_N\}$ are collected by \textit{independent and identical} experiments. Every time the experiment is performed, the system is reset to some specific known initial state $\rho_0$. (b) The sequential scheme for estimating a parameter $g$ of a quantum system, where the measurements are performed sequentially to collect data $\{s_1,\ldots,s_N\}$ \textit{without resetting the state of the system every after the measurement} and the initial state $\rho_0$ can be arbitrary.}
\label{fig:Schemes}
\end{figure}

In the standard approach one typically focuses on the case where several (say $N$) identical copies of $\rho_g$ are available to experimentalists, who can hence rely on the statistical inference extracted from independent and identically distributed (i.i.d.) measurement outcomes to estimate the value of $g$. 
This scenario is particularly well formulated by those configurations where the unknown parameter $g$ is  associated with some black-box  transformation $\Lambda_g$ (say a phase shift induced in one arm of an interferometric setup) which acts on the input state $\rho_0$ of a  probing system (say the light beam injected into the interferometer) yielding $\rho_g=\Lambda_g(\rho_0)$ as the output density matrix to be measured, with such test repeated $N$ times to collect data $\{s_1,\ldots,s_N\}$. 
See Fig.\ \ref{fig:Schemes}(a). 
In this context, the ultimate limits on the attainable precision in the estimation of $g$, optimized with respect to the general detection strategy, can be computed, resulting in the so-called quantum Cram\'{e}r-Rao bound, which exhibits the functional dependence upon $\rho_g$ via the quantum Fisher information. 
See e.g.\ Refs.\ \cite{ref:QuantumMetrologyVittorio,ref:MetrologyNaturePhoto,ref:Helstrom,ref:BraunsteinCave1994,ref:BraunsteinCave1996AnnPhys,ref:QuantumEstimation,ref:Paris-IJQI,ref:HolevoSNS}.

In many situations of physical interest, however, the possibility of reinitializing the setup to the same state is not necessarily guaranteed.
In the present study we are going to consider a different scheme, in which a
single probing system undergoes multiple applications of $\Lambda_g$ while being monitored during the process without being reinitialized to the same input state.
See Fig.\ \ref{fig:Schemes}(b).
The data $\{s_1,\ldots,s_N\}$ collected by such sequential measurements will be non-i.i.d.\ in general.
Still, we are able to estimate the target parameter $g$ from the data under certain conditions.
We will see that the property of the channel describing the process is important.
The idea is to let the probing system forget about the past by the mixing of the channel \cite{ref:TerhalDiVincenzo,ref:Mixing-Wolf,ref:MixingNJP-BurgarthGiovannetti,ref:ConvexErgodicity} (the channel being intrinsically mixing or designed to be mixing), which clusters the data and allows the central limit theorem to hold for appropriately chosen functionals of the data.
This sequential scheme is suited to account for estimation procedures where one aims to recover $g$ via a sequence of weak measurements that slightly perturb the probe. 
In particular it can be adapted to study physical setups where the probing system is a proper subset of a many-body quantum system which is directly affected by the black-box generator $\Lambda_g$ (an explicit example of this scenario will be analyzed in the final part of this paper).

Various schemes for quantum parameter estimation based on repetitive or continuous measurements have been studied: see e.g.\ \cite{ref:Mabuchi-QSO1996,ref:GambettaWiseman-PRA2001,ref:Guta-PRA,ref:Guta-ContTime,ref:Molmer-PRA2013,ref:Molmer-PRL2014,ref:RybarZiman,ref:GutaHorssen-JMP2015,ref:ZenoEst}.
Among them, analogous setups were analyzed in Refs.\ \cite{ref:Guta-PRA,ref:GutaHorssen-JMP2015}, where the problem was formalized in terms of quantum Markov chains.
Specifically in Ref.\ \cite{ref:Guta-PRA} it has been shown that, under rather general assumptions, the statistics of the associated estimation problem converges asymptotically to a normal one, generalizing the similar results which were known to apply to purely classical settings \cite{HOPFNER,ref:DasGuptaCh10,ref:BillingsleyTh27.4}. 
In the present paper we first provide an independent derivation of the previous result \cite{ref:Guta-PRA} via a diagrammatic approach to compute the leading-order contributions to the moments of the associated estimating functional of the data $\{s_1,\ldots,s_N\}$, i.e., the moments of the average $S=\frac{1}{N}\sum_{i=1}^Ns_i$.
This approach allows us to prove the central limit theorem including other estimating functionals capturing the correlations among different measurement outcomes, e.g., $C_\ell=\frac{1}{N-\ell}\sum_{i=1}^{N-\ell}s_is_{i+\ell}$.
The asymptotic normality of the empirical measure associated to chains of subsequent measurement outcomes is proved in Ref.\ \cite{ref:GutaHorssen-JMP2015}, but in contrast to this previous work we provide explicit formulas which allow us to evaluate the elements of the covariance matrix of the normal distribution of the variables $S$ and $C_\ell$.
Moreover, we point out that the inclusion of the correlations $C_\ell$ for estimation, which do not contain any useful information in the usual i.i.d.\ data, can help improve the accuracy of the estimation.
This result, while not conclusive, is a preliminary (yet nontrivial) step towards the determination of the ultimate accuracy limit attainable in the non-i.i.d.\ settings.

This paper is organized as follows. 
In Sec.\ \ref{sec:ErgodicMixing} we introduce the notation and recall some basic mathematical facts which will be used in the paper. 
The non-i.i.d.\ estimation model is then presented in Sec.\ \ref{sec:Scheme}\@.
In Sec.\ \ref{sec:CLTS} we focus on the simplest estimating functional $S$ of the measurement data, and prove its asymptotic normality under the assumption of mixing of the process.
The central limit theorem is generalized to include the correlations $C_\ell$ and their role in the estimation problem is addressed in Sec.\ \ref{sec:Correlation}\@. 
An explicit example is then presented in Sec.\ \ref{sec:ex}, where we discuss the estimation of the temperature of a thermal reservoir via local measurements on a quantum probe in contact with the reservoir.
Conclusions and perspectives are summarized in Sec.\ \ref{sec:conc}, while some technical elements are presented in the Appendices.

\section{Notation and Mathematical Background}
\label{sec:ErgodicMixing}
In this section we introduce the notation and recall some basics facts on the theory of quantum channels.

\subsection{Quantum Ergodic/Mixing Channels}
Quantum channels are completely positive and trace-preserving (CPTP) maps, transforming density operators to density operators \cite{ref:NielsenChuang,ref:DynamicalMap-Alicki,ref:HolevoVittorioReview}.
Every CPTP map $\mathcal{E}$ admits at least one \textit{fixed point} \cite{ref:TerhalDiVincenzo,ref:Mixing-Wolf}, namely, a stationary state $\rho_*$,
\begin{equation}
\mathcal{E}(\rho_*)=\rho_*,
\end{equation}
which is hermitian, positive-semidefinite, and of unit trace.
In other words, the fixed point $\rho_*$ is an eigenstate of the map $\mathcal{E}$ belonging to its unit eigenvalue $1$.

If the fixed point $\rho_*$ is unique, the quantum channel $\mathcal{E}$ is called \textit{ergodic} \cite{ref:Mixing-Wolf,ref:MixingNJP-BurgarthGiovannetti,ref:ConvexErgodicity}.
It implies
\begin{equation}
\frac{1}{N}\sum_{n=0}^{N-1}\mathcal{E}^n(\rho_0)
\xrightarrow{N\to\infty}
\rho_*,\quad
\forall\ \text{states}\ \rho_0,
\label{eqn:Ergodic}
\end{equation}
where $\mathcal{E}^n=\mathcal{E}\circ\cdots\circ\mathcal{E}$ denotes $n$ recursive applications of the channel $\mathcal{E}$, and the convergence is in the superoperator norm with corrections 
whose leading order scales as $1/N$. 
Moreover, if the fixed point $\rho_*$ is unique and the unit eigenvalue $1$ is the only \textit{peripheral eigenvalue} (eigenvalue of unit magnitude), the quantum channel $\mathcal{E}$ is converging as
\begin{equation}
\mathcal{E}^N(\rho_0)
\xrightarrow{N\to\infty}
\rho_*,\quad
\forall\ \text{states}\ \rho_0,
\label{eqn:Mixing}
\end{equation}
and is called \textit{mixing}, with the convergence being as in (\ref{eqn:Ergodic}) \cite{ref:TerhalDiVincenzo,ref:Mixing-Wolf,ref:MixingNJP-BurgarthGiovannetti,ref:ConvexErgodicity}.
Mixing implies ergodicity, but the converse is not necessarily true.

As commented above, the fixed point $\rho_*$ of a quantum channel $\mathcal{E}$ is an eigenstate of $\mathcal{E}$ belonging to its unit eigenvalue $1$.
In a matrix representation of $\mathcal{E}$, it is a ``right eigenvector.''
The corresponding ``left eigenvector'' belonging to the same eigenvalue can be different from the right eigenvector in general.
For a quantum channel $\mathcal{E}$, the trace $\Tr$ is a left eigenvector belonging to the unit eigenvalue $1$, since the quantum channel $\mathcal{E}$ is trace-preserving,
\begin{equation}
\Tr\{\mathcal{E}(\rho)\}=\Tr\rho.
\label{eqn:TP}
\end{equation}
Let us hence write the fixed point $\rho_*$ and the trace $\Tr$ in the vectorized notation as
\begin{equation}
\rho_*\leftrightarrow|\rho_*),\qquad
\Tr\leftrightarrow(1|,
\end{equation}
respectively, and a couple of eigenvalue equations for the unit eigenvalue $1$ read
\begin{equation}
\mathcal{E}|\rho_*)=|\rho_*),\qquad
(1|\mathcal{E}=(1|.
\label{eqn:Eigenvectors1}
\end{equation}

More explicitly, given any complete set of orthonormal basis states $\{\ket{n}\}_n$ of the system, an operator $A=\sum_{n,n'}A_{nn'}\ket{n}\bra{n'}$ is vectorized by $|A)=\sum_{n,n'}A_{nn'}\ket{n}\otimes\ket{n'}$ \cite{ref:GeometryOfQuantumStates}.
The trace $(1|=\sum_n\bra{n}\otimes\bra{n}$ is (the hermitian conjugate of) the vectorized version of the identity operator: that is why it is denoted by $(1|$.
In addition, the inner product $(A|B)=\Tr\{A^\dag B\}$ is the Hilbert-Schmidt inner product.
In this representation, the quantum channel $\mathcal{E}\leftrightarrow\sum_{m,n,m',n'}\mathcal{E}_{mn,m'n'}\ket{m}\bra{m'}\otimes\ket{n}\bra{n'}$ is a matrix with the matrix elements $\mathcal{E}_{mn,m'n'}=\bra{m}\mathcal{E}(\ket{m'}\bra{n'})\ket{n}$ in the original representation, and the application of a map $\mathcal{E}$ is expressed by the multiplication of the corresponding matrix.
By abuse of notation we use the same symbol $\mathcal{E}$ for its matrix representation.

In this matrix representation, the eigenvalue equation for $\mathcal{E}$ reads
\begin{equation}
\mathcal{E}|u_n)=\lambda_n|u_n),\qquad
(v_n|\mathcal{E}=\lambda_n(v_n|.
\end{equation}
In particular, $|u_0)=|\rho_*)$ and $(v_0|=(1|$ with $\lambda_0=1$.
The eigenvectors belonging to different eigenvalues are orthogonal to each other and normalized as
\begin{equation}
(v_n|u_n)=1,\qquad
(v_m|u_n)=0\ \ \text{for}\ \ \lambda_m\neq\lambda_n.
\end{equation}
Note that the matrix $\mathcal{E}$ might not be diagonalizable but is cast in the Jordan canonical form in general \cite{ref:TerhalDiVincenzo,ref:Mixing-Wolf}.

In this paper, ergodic or mixing channels will play a central role.
The unit eigenvalue $1$ of such a channel $\mathcal{E}$ is not degenerated by definition, and the ergodic/mixing channel $\mathcal{E}$ can always be decomposed as
\begin{equation}
\mathcal{E}
=\mathcal{P}_*+\mathcal{E}',
\label{eqn:Decomp}
\end{equation}
where 
\begin{equation}
\mathcal{P}_*=|\rho_*)(1|=\rho_*\Tr\{{}\bullet{}\}
\label{eqn:Pfixed}
\end{equation}
is the eigenprojection belonging to the nondegenerate unit eigenvalue $1$ of $\mathcal{E}$, and the remaining part $\mathcal{E}'$ (which is not CPTP) is built on the eigenvectors $\{|u_n)\}_{n\neq0}$ and $\{(v_n|\}_{n\neq0}$ belonging to the eigenvalues $\lambda_n$ different from $1$.
By construction $\mathcal{E}'$ is orthogonal to $\mathcal{P}_*$, i.e., 
$
\mathcal{P}_*\mathcal{E}'
=\mathcal{E}'\mathcal{P}_*=0
$.
Moreover, since $\mathcal{E}'$ does not admit a unit eigenvalue $1$, the inverse $(1-\mathcal{E}')^{-1}$ exists, and we have
\begin{equation}
\frac{1}{N}\sum_{n=0}^{N-1}\mathcal{E}^n
=\mathcal{P}_*+\frac{1}{N}\frac{1-\mathcal{E}'^N}{1-\mathcal{E}'}\mathcal{Q}_*
\label{eqn:GeometricSum}
\end{equation}
with 
$
\mathcal{Q}_*=1-\mathcal{P}_*
$,
which allows us to prove the convergence to the fixed point $\rho_*$ in (\ref{eqn:Ergodic}).
If the channel $\mathcal{E}$ is not only ergodic but also mixing, the spectral radius of $\mathcal{E}'$ is strictly smaller than $1$, and we get
\begin{equation}
\mathcal{E}^N
=\mathcal{P}_*+\mathcal{E}'^N
\xrightarrow{N\to\infty}\mathcal{P}_*,
\end{equation}
which proves (\ref{eqn:Mixing}).
If the channel $\mathcal{E}$ is ergodic but not mixing, $\mathcal{E}'$ admits a peripheral eigenvalue, and $\mathcal{E}'^N$ does not decay: we lose the convergence (\ref{eqn:Mixing}), but the averaged channel converges as (\ref{eqn:Ergodic}).

Note again that $\mathcal{E}'$ might not be diagonalizable if some of the eigenvalues $\lambda_n$ of $\mathcal{E}$ are degenerated, but it is not a problem for the convergence: see \cite{ref:TerhalDiVincenzo,ref:Mixing-Wolf,ref:MixingNJP-BurgarthGiovannetti,ref:ConvexErgodicity}.

\subsection{Measurement and Back-Action}  
We recall that in quantum mechanics the most general detection scheme can be formalized in terms of positive operator-valued measure (POVM). 
See e.g.\ Ref.\ \cite{ref:NielsenChuang}. 
Expressed in the superoperator language this accounts to assigning a collection $\mathfrak{M}=\{\mathcal{M}_s\}_s$ of trace-decreasing channels $\mathcal{M}_s$ describing the statistics of the measurement and the back-action on the probed system.
In particular, given $\rho$ the density matrix of the system before the measurement, the probability of getting outcome $s$ by the measurement $\mathfrak{M}$ is given by
\begin{equation}\label{neweq} 
p(s|\rho)  = \Tr \{ \mathcal{M}_s (\rho) \} = (1| \mathcal{M}_s |\rho),
\end{equation}
with $\mathcal{M}_s(\rho)$ being the conditional (not normalized) state immediately after the event.
By construction the map 
\begin{equation}
\mathcal{M} = \sum_s \mathcal{M}_s,
\label{EMMEMEDIA}
\end{equation} 
obtained by summing over all possible values of $s$, is CPTP and describes the evolution of the system when no record of the measurement outcome is kept.
We also notice that given $\mathcal{D}$ and $\mathcal{E}$ two CPTP maps, the set of channels $\mathcal{M}_s'= \mathcal{E}\mathcal{M}_s\mathcal{D}$ defines a new POVM measurement $\mathfrak{M}'=\{\mathcal{M}_s'\}_s$, where immediately before and after the measurement $\mathfrak{M}$ one transforms the state of the probed system through the actions of $\mathcal{D}$ and $\mathcal{E}$, respectively. 
Finally we observe that given $\mathfrak{M} = \{ \mathcal{M}_r\}_r$ and $\mathfrak{N} = \{\mathcal{N}_s\}_s$ two POVMs, the operator $(\mathcal{N}_s \circ \mathcal{M}_r )(\rho)$ represents the conditional (not normalized) state obtained when the measurements $\mathfrak{M}$ and $\mathfrak{N}$ are performed on a system in the state $\rho$ yielding measurement outcomes $r$ and $s$, respectively, the associated probability given by $p(r,s|\rho) =(1|\mathcal{N}_s \mathcal{M}_r |\rho)$.

\section{Sequential Scheme}
\label{sec:Scheme}
The problem we study is the following: we wish to recover an unknown parameter $g$ of a quantum system, which is encoded in the state of a quantum probe via the
action of a quantum channel $\Lambda_g$, 
\begin{equation} 
\rho_0 \mapsto \Lambda_g (\rho_0).
\end{equation}
Here $\rho_0$ is the input state of the probe, which (possibly) is initialized by us, while $\Lambda_g(\rho_0)$ is the associated output state, on which we are allowed to perform measurement in order to learn about $g$. 
In a standard i.i.d.\ approach \cite{ref:QuantumEstimation}, one is supposed to perform the same experiment several times collecting i.i.d.\ outcomes $\{s_1,\ldots,s_N\}$, from which  
the value of $g$ is to be extrapolated via some suitable data processing. 
See Fig.\ \ref{fig:Schemes}(a). 
More precisely, in every experimental run of such an i.i.d.\ scheme the probe should be initialized in the \textit{same} input state $\rho_0$ and the \textit{same} POVM measurement $\mathfrak{M} = \{ \mathcal{M}_s\}_s$ should be performed after $\Lambda_g$ has operated on the probe. 
On the contrary, in the protocol we are going to discuss here, while we keep performing the same measurement $\mathfrak{M}$ on the probe, the probe is not reset to $\rho_0$ after each measurement step. 
Instead, we just repeat the application of $\Lambda_g$ followed by a measurement many times to get a sequence of  outcomes $\{s_1,\ldots,s_N\}$, whose statistics is not necessarily i.i.d.\ anymore. 
See Fig.\ \ref{fig:Schemes}(b).
In this scenario, following the framework detailed in Sec.\ \ref{sec:ErgodicMixing}, the state of the probe undergoes a conditional evolution described by the (not necessarily normalized) density matrix 
\begin{equation}
\rho_0\mapsto(\mathcal{E}_{s_N}\circ{}\cdots{}\circ\mathcal{E}_{s_1})(\rho_0),
\quad
\mathcal{E}_s=\mathcal{M}_s\circ \Lambda_g,
\label{eqn:Evo}
\end{equation}
whose trace
\begin{equation}\label{probseq}
p(s_1,\ldots,s_N|\rho_0)
=(1|\mathcal{E}_{s_N}\cdots{}\mathcal{E}_{s_1}|\rho_0)
\end{equation}
defines the probability of the associated measurement event.

It is worth observing that this mathematical setting includes the i.i.d.\ scenario as a special case, where $\Lambda_g$ is identified with $\Lambda_g \circ \mathcal{P}_0$, with $\mathcal{P}_0=|\rho_0)(1|=\rho_0\Tr\{{}\bullet{}\}$ being the map resetting the state of the probe into $\rho_0$.
Indeed, with this choice the probability (\ref{probseq}) coincides with the one for the case where the measurements $\mathfrak{M} = \{ \mathcal{M}_s\}_s$ are performed independently on $N$ copies of $\Lambda_g(\rho_0)$, i.e.,
\begin{align}
p(s_1,\ldots,s_N|\rho_0)
&
=(1|\mathcal{E}_{s_N}|\rho_0)\cdots(1| \mathcal{E}_{s_2}|\rho_0)(1|\mathcal{E}_{s_1}|\rho_0)\nonumber \\
&
=p(s_N|\rho_0)\cdots p(s_2|\rho_0)p(s_1|\rho_0).
\label{probseq1}
\end{align}

From (\ref{EMMEMEDIA}) it follows that the map $\mathcal{E}$ obtained by summing $\mathcal{E}_s$ in (\ref{eqn:Evo}) over all $s$, 
\begin{equation}
\mathcal{E}=\sum_s\mathcal{E}_s = {\cal M} \circ \Lambda_g, 
\label{eqn:E}
\end{equation}
is CPTP, ensuring the proper normalization of the probability (\ref{probseq}).
As an additional constraint we will require it to be mixing (in some cases, e.g., in Sec.\ \ref{sec:LARGE}, however, we will weaken this requirement by imposing $\mathcal{E}$ to be just ergodic).
This is not a strong assumption, as mixing channels actually form an open and dense set. Under this condition we will be able to prove that the parameter $g$ can be estimated from the single sequence of data $\{s_1,\ldots,s_N\}$ collected by the sequential measurements, irrespective of the initial state $\rho_0$ \cite{ref:Guta-PRA}.
The rough idea is that, thanks to the mixing (\ref{eqn:Mixing}), repeated applications of the channel force the quantum system to forget its initial state and at the same time decorrelate the data separated beyond the correlation length, which clusters the data and allows us to define \textit{self-averaging} quantities as estimating functionals, whose fluctuations diminish as $N$ increases, i.e., the central limit theorem holds.

\subsection*{Inferring $\bm{g}$ from $\bm{\{ s_1, \ldots, s_N\}}$} 
A standard estimating functional of the measured data $\{s_1,\ldots,s_N\}$, through which one tries to infer the value of $g$, is the average
\begin{equation}
S=\frac{1}{N}\sum_{i=1}^Ns_i.
\label{eqn:S}
\end{equation}
In Ref.\ \cite{ref:Guta-PRA} it was noted that under the assumption that the average channel $\mathcal{E}$ in (\ref{eqn:E}) is mixing the central limit theorem holds for $S$, and for large $N$ the probability distribution $P(S)$ of $S$ asymptotically becomes a Gaussian peaked at a value $\langle S\rangle_*$ with a shrinking variance $\sigma^2/N$, which are both independent of the input state $\rho_0$ of the probe, i.e., 
\begin{equation} \label{ASYMPDIS} 
P(S)\simeq\frac{1}{\sqrt{2\pi\sigma^2/N}}e^{-\frac{(S-\langle S\rangle_*)^2}{2\sigma^2/N}}.
\end{equation}
The explicit expressions for $\langle S\rangle_*$ and  $\sigma$ will be provided in (\ref{eqn:Sst}) and (\ref{eqn:sigma}), respectively, in Sec.\ \ref{sec:CLTS} below. 
This ensures that the quantity $S$ evaluated from the single sequence of measurement outcomes is expected, with a high probability, to be very close to its expectation value $\langle S\rangle_*$ with a vanishingly small variance $\sigma^2/N$ for large $N$.
Therefore, by comparing the observed value of $S$ with the formula for the expectation value $\langle S\rangle_*$ as a function of $g$, one can infer the parameter $g$.
It is worth stressing once more that in the sequential scheme the measurement data are not independent of each other.
Therefore, it is not trivial whether the central limit theorem holds, which is usually based on i.i.d.\ data set.
The mixing, however, is strong enough to kill the correlations between two data if they are sufficiently far away from each other, and clusters the data, allowing the central limit theorem to hold.

Thanks to (\ref{ASYMPDIS}) the uncertainty in the estimation of  $g$ through the quantity $S$ can be evaluated via the Cram\'er-Rao bound as \cite{ref:Helstrom,ref:BraunsteinCave1994,ref:BraunsteinCave1996AnnPhys,ref:QuantumEstimation,ref:Paris-IJQI,ref:HolevoSNS,ref:MetrologyNaturePhoto,ref:Cramer}
\begin{equation}
\delta g\simeq\frac{1}{\sqrt{\mathcal{F}(g)}},
\label{eqn:CR}
\end{equation}
where $\mathcal{F}(g)$ is the Fisher information of the problem given by
\begin{equation}
\mathcal{F}(g)
=\int dS\,P(S)\left(\frac{\partial}{\partial g}\ln P(S)\right)^2
\simeq\frac{N}{\sigma^2}\left(\frac{\partial\langle S\rangle_*}{\partial g}\right)^2.
\label{eqn:FS}
\end{equation}
Accordingly, as long as $\langle S\rangle_*$ exhibits a nontrivial functional dependence upon $g$, the Fisher information $\mathcal{F}(g)$ increases linearly in $N$, yielding an estimation error (\ref{eqn:CR}) which diminishes as $\delta g\simeq1/\sqrt{N}$ [in (\ref{eqn:FS}) we have omitted the contribution from $\partial\sigma/\partial g$ to the Fisher information $\mathcal{F}(g)$ since it does not grow with $N$].

It may happen however that  the quantity $\langle S\rangle_*$ does not depend upon $g$.
In such a case $\mathcal{F}(g)$ nullifies, signaling that  it is impossible to recover $g$ through $S$ [a problem which cannot be fixed by properly choosing the input state $\rho_0$ of the probe, the asymptotic distribution (\ref{ASYMPDIS}) being independent of $\rho_0$]. 
Nonetheless, even in this particular case, the sequence of data $\{s_1,\ldots,s_N\}$, which is not i.i.d.\ in general, can still contain some functional dependence upon $g$, which can be exploited for the estimation of $g$. 
In particular, the aim of the present work is to show that the correlations among the measurement data, which are absent in the usual i.i.d.\ data, can be used for this purpose.
It turns out that, under the same mixing assumption on the channel $\mathcal{E}$ that leads to the central limit theorem for $S$ in (\ref{ASYMPDIS}), the correlations are also self-averaging and become asymptotically normal for large $N$, enabling one to estimate $g$ through them.
See (\ref{eqn:PX}) and (\ref{eqn:FSC}) in Sec.\ \ref{sec:Correlation} below. 
Even in the case where $\langle S\rangle_*$ depends upon $g$, looking also at the correlations help enhance the precision of the estimation of $g$, which will be demonstrated in Fig.\ \ref{fig:Fcomp} with the example studied in Sec.\ \ref{sec:ex}\@.

We first present an alternative derivation of the results of Ref.\ \cite{ref:Guta-PRA}, i.e., the asymptotic normality of $S$, on the basis of a diagrammatic approach in Sec.\ \ref{sec:CLTS}\@. While our approach is more involved than the elegant perturbative approach taken in Ref.\ \cite{ref:Guta-PRA}, it allows us to generalize the scheme to include the correlations among the measurement data in a straightforward manner to enhance the precision of the estimation. We shall indeed prove the asymptotic normality of variables including the correlation functionals in Sec.\ \ref{sec:Correlation}\@.

\section{Statistical Behavior of $\bm{S}$}
\label{sec:CLTS}
This section is devoted to provide an alternative derivation of the results of Ref.\ \cite{ref:Guta-PRA},  which ultimately leads to the asymptotic normality of $S$ in (\ref{ASYMPDIS}).
We start in Sec.\ \ref{sec:LARGE} by proving that under the hypothesis that the average channel $\mathcal{E}$ in (\ref{eqn:E}) is ergodic the quantity $S$ is self-averaging, converging to a fixed value $\langle S\rangle_*$ independent of the input state $\rho_0$. 
Then in Sec.\ \ref{sec:central} we introduce the mixing property and show that under this stronger condition the distribution $P(S)$, which rules the statistics of $S$, becomes asymptotically normal.

\subsection{Law of Large Numbers by Ergodicity}\label{sec:LARGE} 
Consider the expectation value of the quantity $S$ with respect to the probability (\ref{probseq}) governing the statistical distribution of the measurement outcomes, i.e., 
\begin{align}
\langle S\rangle_N
&=\sum_{s_1}\cdots\sum_{s_N}S\,p(s_1,\ldots,s_N|\rho_0)
\nonumber\displaybreak[0]\\
&=\frac{1}{N}\sum_{i=1}^N\sum_{s_1}\cdots\sum_{s_N}
s_i(1|\mathcal{E}_{s_N}\cdots\mathcal{E}_{s_1}|\rho_0)
\nonumber\displaybreak[0]\\
&=\frac{1}{N}\sum_{i=1}^N
(1|\mathcal{E}^{(1)}\mathcal{E}^{i-1}|\rho_0),
\label{eqn:ExpS}
\end{align}
where $\mathcal{E}^{(1)}$ is defined by
\begin{equation}
\mathcal{E}^{(1)}=\sum_ss\,\mathcal{E}_s.
\end{equation}
Assume here that the channel $\mathcal{E}$
is ergodic with unique fixed point $\rho_*$: using (\ref{eqn:GeometricSum}), the right-hand side of (\ref{eqn:ExpS}) can be written as  
\begin{equation}
\langle S\rangle_N
=\langle S\rangle_*
+\frac{1}{N}
(1|
\mathcal{E}^{(1)}\frac{1-\mathcal{E}'^N}{1-\mathcal{E}'}
\mathcal{Q}_*
|\rho_0).
\label{eqn:Slimit}
\end{equation}
The first contribution 
\begin{equation}
\langle S\rangle_*
=(1|\mathcal{E}^{(1)}|\rho_*)
=\sum_s s\Tr\{ {\cal E}_s (\rho_*)\} =  \langle s\rangle_*
\label{eqn:Sst}
\end{equation}
is the value of $\langle S\rangle_N$ when the input state $\rho_0$ of the probe coincides with the fixed point $\rho_*$ of $\mathcal{E}$.
As stressed by the last identity, it also coincides with the expectation value associated with the i.i.d.\ measurement on $\rho_*$ with the POVM $\mathfrak{M}$. 
The second contribution in (\ref{eqn:Slimit}) instead is a correction which scales at most as $1/N$ for any other choice of $\rho_0$. 
Accordingly in the large-$N$ limit we get
\begin{equation}
\langle S\rangle_N \xrightarrow{N\to\infty} \langle S\rangle_*,
\label{eqn:SlimitLIMIT}
\end{equation}
irrespectively of $\rho_0$.

In a similar way we can compute the variance of $S$, obtaining 
\begin{align}
(\Delta S)_N^2
={}&
\langle S^2\rangle_N
-\langle S\rangle_N^2
\nonumber\\
={}&
\frac{1}{N}
(\Delta s)_*^2
+\frac{2}{N}
(1|
\tilde{\mathcal{E}}^{(1)}\frac{\mathcal{Q}_*}{1-\mathcal{E}'}\tilde{\mathcal{E}}^{(1)}
|\rho_*)
\nonumber\displaybreak[0]\\
&{}
-\frac{2}{N^2}
\sum_{j=1}^{N-1}
(1|
\tilde{\mathcal{E}}^{(1)}\frac{\mathcal{E}'^{N-j}}{1-\mathcal{E}'}\mathcal{Q}_*\tilde{\mathcal{E}}^{(1)}\mathcal{E}'^{j-1}\mathcal{Q}_*
|\rho_0)
\nonumber\displaybreak[0]\\
&{}+O(1/N^2),
\label{eqn:VarS}
\end{align}
where $\langle S^2\rangle_N$ is defined similarly to $\langle S\rangle_N$ in (\ref{eqn:ExpS}), and
\begin{equation}
\tilde{\mathcal{E}}^{(m)}=\sum_s(\delta s)^m\mathcal{E}_s,\qquad
\delta s=s-\langle s\rangle_*,
\label{eqn:En}
\end{equation}
while
\begin{equation} 
(\Delta s)^2_*=\langle s^2\rangle_*-\langle s\rangle_*^2=(1|\tilde{\mathcal{E}}^{(2)}|\rho_*)
\label{eqn:DeltasStar}
\end{equation} 
is the variance of $s$ in the stationary state $\rho_*$ as in (\ref{eqn:Sst}).
Equation (\ref{eqn:VarS}) shows that the variance shrinks as $(\Delta S)_N^2\sim1/N$, and the fluctuation of $S$ around $\langle S\rangle_N$ becomes smaller and smaller as we proceed with the measurements.
As a result, the probability of finding a single-shot value $S$ close to its expectation value $\langle S\rangle_N$ becomes very high.
Indeed, Chebyshev's inequality bounds the probability of $S$ deviating from $\langle S\rangle_N$ as
\begin{equation}
\mathop{\text{Prob}}\Bigl(
|S-\langle S\rangle_N|>K(\Delta S)_N
\Bigr)<\frac{1}{K^2}
\end{equation}
for any positive $K$. 
In this way, $S$ is self-averaging: each single $S$ is very close to its expectation value with very high probability.
In addition, as shown in (\ref{eqn:SlimitLIMIT}), $\langle S\rangle_N$ becomes independent of the initial state $\rho_0$.

\subsection{Beyond the Law of Large Numbers:\\Central Limit Theorem by Mixing}\label{sec:central}
We have so far assumed the ergodicity of the channel $\mathcal{E}$: this is necessary and sufficient for the convergence $\langle S\rangle_N\to\langle S\rangle_*$ in (\ref{eqn:SlimitLIMIT}), and for the shrinking variance $(\Delta S)_N^2\sim1/N$ in (\ref{eqn:VarS}).
If we further assume that $\mathcal{E}$ is mixing, we can say more.
For instance, the third contribution to the variance $(\Delta S)_N^2$ in (\ref{eqn:VarS}) decays as $\mathcal{E}'^N/N$ [note that the sum over $j$ accumulates to $O(N)$], i.e., faster than $1/N$ (it is not guaranteed under the ergodicity, since $\mathcal{E}'^N$ does not decay), and the variance $(\Delta S)_N^2$ asymptotically becomes independent of the initial state $\rho_0$.
This is because the mixing makes the system forget the initial state $\rho_0$ as (\ref{eqn:Mixing}) without averaging along the time trace.

Most importantly, if $\mathcal{E}$ is mixing, one can prove that the probability distribution of $S$ asymptotically becomes normal, converging to the Gaussian distribution (\ref{ASYMPDIS}).  
The asymptotic normality  of $S$ under the mixing condition was proved in Ref.\ \cite{ref:Guta-PRA}.
Here we derive  the same result by introducing a diagrammatic approach. 
Specifically, in the following subsection we shall compute the moments of the variable $S-\langle S\rangle_*$, showing that for large $N$ they admit the scaling 
\begin{equation}
\langle(S-\langle S\rangle_*)^n\rangle_N\sim O(1/N^{\lceil\frac{n}{2}\rceil}),
\label{eqn:Scaling}
\end{equation}
where $\lceil x\rceil$ denotes the smallest integer not less than $x$.  
In particular for even $n$ we shall see that 
the leading-order term is given by
\begin{equation}
\langle(S-\langle S\rangle_*)^n\rangle_N
=\frac{n!}{(2N)^{n/2}(n/2)!}\sigma^n
+O(1/N^{n/2+1})
\label{eqn:EvenMoments}
\end{equation}
with 
\begin{equation}
\sigma^2
=(1|\tilde{\mathcal{E}}^{(2)}|\rho_*)
+2(1|\tilde{\mathcal{E}}^{(1)}\frac{\mathcal{Q}_*}{1-\mathcal{E}'}\tilde{\mathcal{E}}^{(1)}|\rho_*),
\label{eqn:sigma}
\end{equation}
where $\tilde{\mathcal{E}}^{(1)}$ and $\tilde{\mathcal{E}}^{(2)}$ are defined as in (\ref{eqn:En}).
These results allow us to conclude that the characteristic function for the scaled variable $x=\sqrt{N}(S-\langle S\rangle_*)$ 
becomes asymptotically normal in the limit $N\to\infty$. Indeed by direct substitution we have
\begin{align}
\chi(k)
&=\langle e^{ik\sqrt{N}(S-\langle S\rangle_*)}\rangle_N
\nonumber\\
&=\sum_{n=0}^\infty\frac{1}{n!}(ik\sqrt{N})^n\langle(S-\langle S\rangle_*)^n\rangle_N
\nonumber\\
&\xrightarrow{N\to\infty}\sum_{r=0}^\infty
\frac{1}{2^rr!}(ik\sigma)^{2r}
=e^{-\frac{1}{2}k^2\sigma^2}.
\label{eqn:CharFuncScaled}
\end{align}
Accordingly the central limit theorem holds and in the limit $N\to\infty$ the probability distribution $P(x)$ of $x$ converges to a Gaussian peaked at $x=0$ with variance $\sigma^2$, i.e., 
$P(x)\to e^{-x^2/2\sigma^2}/\sqrt{2\pi\sigma^2}$,
which in the original variable implies (\ref{ASYMPDIS}).

\subsubsection*{Diagrammatic Approach to Evaluate the Moments of $S$}
The expression for the first moment of $S-\langle S\rangle_*$ follows from  (\ref{eqn:Slimit}) and is equal to 
\begin{equation}
\langle S-\langle S\rangle_*\rangle_N
=\frac{1}{N}
(1|
\tilde{\mathcal{E}}^{(1)}\frac{1-\mathcal{E}'^N}{1-\mathcal{E}'}
\mathcal{Q}_*
|\rho_0),
\label{eqn:1stMoment}
\end{equation}
which scales as $1/N$ as anticipated. 
Analogously the second moment is readily obtained from (\ref{eqn:VarS}) by noticing that 
\begin{equation}
(\Delta S)_N^2
=\langle(S-\langle S\rangle_*)^2\rangle_N-(\langle S\rangle_N-\langle S\rangle_*)^2.
\end{equation}
For future reference we find it useful to rederive it:
\begin{align}
&\langle(S-\langle S\rangle_*)^2\rangle_N
\nonumber\displaybreak[0]\\
&\quad
=\sum_{s_1}\cdots\sum_{s_N}(S-\langle S\rangle_*)^2
p(s_1,\ldots,s_N|\rho_0)
\nonumber\displaybreak[0]\\
&\quad
=\frac{1}{N^2}\sum_{i=1}^N\sum_{j=1}^N
\sum_{s_1}\cdots\sum_{s_N}\delta s_i\,\delta s_j
(1|\mathcal{E}_{s_N}\cdots\mathcal{E}_{s_1}|\rho_0)
\nonumber\displaybreak[0]\\
&\quad
=\frac{1}{N^2}\sum_{i=1}^N
(1|\tilde{\mathcal{E}}^{(2)}\mathcal{E}^{i-1}|\rho_0)
\nonumber\displaybreak[0]\\
&\quad\quad\ %
{}+\frac{2}{N^2}\sum_{j=2}^N\sum_{i=1}^{j-1}
(1|\tilde{\mathcal{E}}^{(1)}\mathcal{E}^{j-i-1}\tilde{\mathcal{E}}^{(1)}\mathcal{E}^{i-1}|\rho_0).
\label{eqn:DeltaS2Def}
\end{align}
To simplify this we insert the decomposition of the ergodic channel $\mathcal{E}$ given in (\ref{eqn:Decomp}), namely, we insert $\mathcal{P}_*=|\rho_*)(1|$ or $\mathcal{E}'$ in place of $\mathcal{E}$.
Notice however that 
\begin{equation}
\langle\delta s\rangle_*
=(1|\tilde{\mathcal{E}}^{(1)}|\rho_*)=0.
\label{eqn:VanishExp}
\end{equation}
Due to this condition, the places in which we can insert $\mathcal{P}_*$ are limited.
The nonvanishing contributions to the second moment hence read
\begin{align}
&\langle(S-\langle S\rangle_*)^2\rangle_N
\nonumber\displaybreak[0]\\
&\ %
=
\frac{1}{N^2}\sum_{i=1}^N
(1|\tilde{\mathcal{E}}^{(2)}|\rho_*)(1|\rho_0)
+\frac{1}{N^2}\sum_{i=1}^N
(1|\tilde{\mathcal{E}}^{(2)}\mathcal{E}'^{i-1}Q_*|\rho_0)
\nonumber\displaybreak[0]\\
&\ \quad
{}+\frac{2}{N^2}\sum_{j=2}^N\sum_{i=1}^{j-1}
(1|\tilde{\mathcal{E}}^{(1)}\mathcal{E}'^{j-i-1}Q_*\tilde{\mathcal{E}}^{(1)}|\rho_*)(1|\rho_0)
\nonumber\displaybreak[0]\\
&\ \quad
{}+\frac{2}{N^2}\sum_{j=2}^N\sum_{i=1}^{j-1}
(1|\tilde{\mathcal{E}}^{(1)}\mathcal{E}'^{j-i-1}Q_*\tilde{\mathcal{E}}^{(1)}\mathcal{E}'^{i-1}Q_*|\rho_0),
\label{eqn:DeltaS2Part}
\end{align}
and the direct computation of the summations yields (\ref{eqn:DeltaS2Complete}) and (\ref{eqn:VarS}).

It is now clear why the second moment $(\Delta S)_N^2$ in (\ref{eqn:VarS}) as well as the second moment $\langle(S-\langle S\rangle_*)^2\rangle_N$ scales as $1/N$.
There are two cases, as we saw in (\ref{eqn:DeltaS2Def}): (i) two points $\delta s_i$ and $\delta s_j$ coincide ($i=j$) and we have a single summation $\frac{1}{N^2}\sum_i$; (ii) two points $\delta s_i$ and $\delta s_j$ do not coincide ($i\neq j$) and we have double summations $\frac{1}{N^2}\sum\sum_{i\neq j}$.
In any case, once $\mathcal{E}^k$ with some power $k$ ($=j-i-1$ or $i-1$ in the above formula for the second moment) is substituted by $\mathcal{P}_*$, a summation accumulates as $\frac{1}{N}\sum\to O(1)$, while the contribution from $\mathcal{E}'$ does not: recall the geometric series in (\ref{eqn:GeometricSum}), where the contribution from $\mathcal{E}'$ remains $O(1/N)$.
Thus, the substitution rules for estimating the scaling are:
\begin{equation}
\frac{1}{N}\sum\mathcal{P}_*\to O(1)
\quad\text{and}\quad
\frac{1}{N}\sum\mathcal{E}'^k\to O(1/N).
\label{eqn:ReplaceRule}
\end{equation}
Due to $(1|\tilde{\mathcal{E}}^{(1)}|\rho_*)=0$ and the coincidence of $\delta s_i$ and $\delta s_j$, we can insert at most one $\mathcal{P}_*$ in place of $\mathcal{E}$ for the second moment $\langle(S-\langle S\rangle_*)^2\rangle_N$: see (\ref{eqn:DeltaS2Part}).
Therefore, the second moment $\langle(S-\langle S\rangle_*)^2\rangle_N$ is at most $O(1/N)$, and so is the variance $(\Delta S)_N^2$.
Note that the second substitution rule in (\ref{eqn:ReplaceRule}) is not valid if the channel $\mathcal{E}$ is ergodic but not mixing.
Indeed, in such a case, the last term in (\ref{eqn:DeltaS2Part}) yields $O(1/N)$, as mentioned in the beginning of this subsection.
The rule is safe if $\mathcal{E}$ is mixing.

We can generalize the above way of estimating the scaling to higher central moments, but a bit more sophisticated rules are required to check the asymptotic normality: we need to care about not only the scalings but also their coefficients.
Anyway, the basic strategy to collect the leading-order contributions is to try to insert $\mathcal{P}_*$ as many times as possible in place of $\mathcal{E}$ avoiding $(1|\tilde{\mathcal{E}}^{(1)}|\rho_*)=0$.
Another important observation is that the insertion of $\mathcal{P}_*=|\rho_*)(1|$ ``breaks'' the process into pieces.
Recognizing these points, we introduce a diagrammatic way of representing the contributions to the moments.

\begin{figure}
\begin{tabular}{l@{\qquad}c}
(a)&\\
&\includegraphics[scale=0.85]{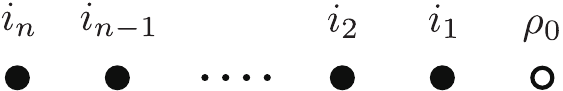}\\[5truemm]
(b)&\\
&\includegraphics[scale=0.75]{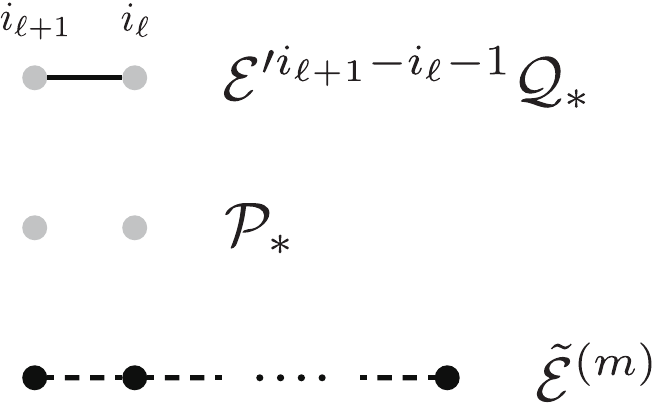}
\end{tabular}
\caption{(a) The $n$ points $\{\delta s_{i_1},\ldots,\delta s_{i_n}\}$ in the $n$th moment $\langle(S-\langle S\rangle_*)^n\rangle_N$ are labeled in chronological order $1\le i_1\le\cdots\le i_N\le N$ and represented by $n$ dots ``$\bullet$'' lined up from right to left.
The right most ``$\circ$'' represents the initial state $|\rho_0)$.
(b) The basic elements for the diagrammatic representation of the contributions to the moments.}
\label{fig:SeqPoints}
\end{figure}

The $n$th moment is given by
\begin{align}
&\langle(S-\langle S\rangle_*)^n\rangle_N
\nonumber\\
&\quad
=\frac{1}{N^n}\sum_{i_1=1}^N\cdots\sum_{i_n=1}^N
\nonumber\\
&\quad\qquad\quad
{}\times\sum_{s_1}\cdots\sum_{s_N}\delta s_{i_1}\cdots\delta s_{i_n}
(1|\mathcal{E}_{s_N}\cdots\mathcal{E}_{s_1}|\rho_0).
\label{eqn:S4Def}
\end{align}
Within the summations over $\{i_1,\ldots,i_n\}$ we relabel the $n$ points $\{\delta s_{i_1},\ldots,\delta s_{i_n}\}$ in chronological order $1\le i_1\le\cdots\le i_N\le N$ and represent them by $n$ dots ``$\bullet$'' lined up in chronological order from right to left.
See Fig.\ \ref{fig:SeqPoints}(a).
The right most ``$\circ$'' represents the initial state $|\rho_0)$, and a trace $(1|$ is supposed to be at the left end.
The points can coincide [$i_{\ell}=i_{\ell+1}$, as in the case $i=j$ for the second moment: see (\ref{eqn:DeltaS2Def})], while between nondegenerate points ($i_{\ell}<i_{\ell+1}$) there are $\mathcal{E}^{i_{\ell+1}-i_\ell-1}$ (with a convention $i_0=0$), which are to be substituted by $\mathcal{P}_*$ or $\mathcal{E}'^{i_{\ell+1}-i_\ell-1}\mathcal{Q}_*$, as we did for the second moment [we need $\mathcal{Q}_*$ to remove $\mathcal{P}_*$ from $\mathcal{E}'^{i_{\ell+1}-i_\ell-1}$ when $i_{\ell+1}-i_\ell-1=0$: see (\ref{eqn:DeltaS2Part})].
\begin{figure}
\includegraphics[width=0.45\textwidth]{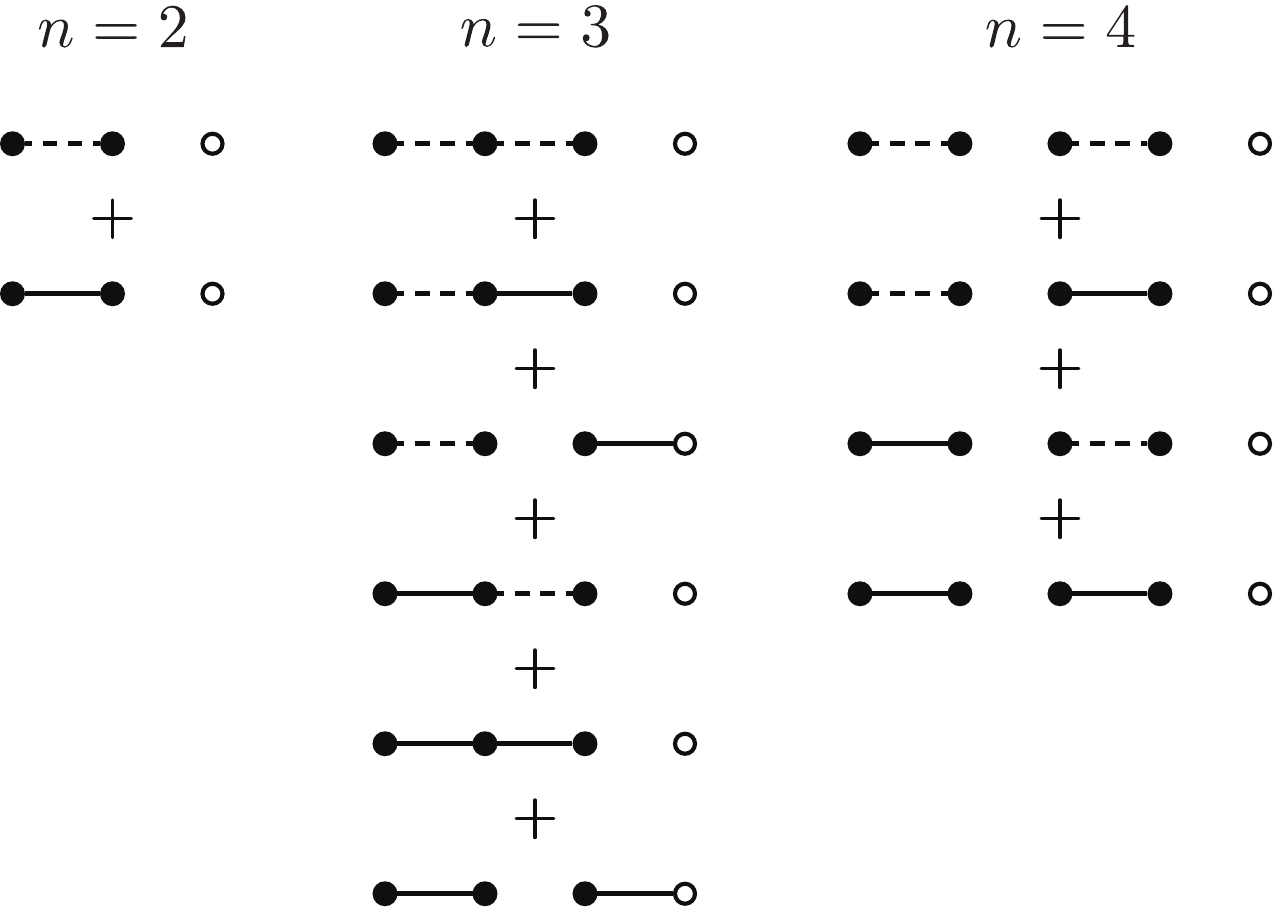}
\caption{The leading-order diagrams for a few lowest  moments $\langle(S-\langle S\rangle_*)^n\rangle_N$.}
\label{fig:Diagrams}
\end{figure}
Now,
\begin{enumerate}
\item[i.] When $\mathcal{E}^{i_{\ell+1}-i_\ell-1}$ between two points is substituted by $\mathcal{E}'^{i_{\ell+1}-i_\ell-1}\mathcal{Q}_*$, we connect the two points by a solid line.
\item[ii.] When $\mathcal{E}^{i_{\ell+1}-i_\ell-1}$ between two points is substituted by $\mathcal{P}_*$, we leave the two points disconnected.
\item[iii.] In the case where two or more points coincide, we connect the points by dashed lines.
(Note that ``$\circ$'' cannot be connected by a dashed line.)
\end{enumerate}
See Fig.\ \ref{fig:SeqPoints}(b).
There are two constraints due to $(1|\tilde{\mathcal{E}}^{(1)}|\rho_*)=0$:
\begin{enumerate}
\item[a.]The left most two points are surely connected either by a solid line or by a dashed line, since we cannot insert $\mathcal{P}_*$ between them due to $(1|\tilde{\mathcal{E}}^{(1)}|\rho_*)=0$ with the left most trace $(1|$.
\item[b.]Each point (except for ``$\circ$'') must be connected with at least one adjacent point either by a solid line or by a dashed line, since we cannot insert $\mathcal{P}_*$ on both sides of a point due to $(1|\tilde{\mathcal{E}}^{(1)}|\rho_*)=0$.
\end{enumerate}
Then, it is easy to draw the diagrams relevant to the leading-order contributions, with the largest possible number of $\mathcal{P}_*$ inserted.
The relevant diagrams for $n=2,3,4$ are shown in Fig.\ \ref{fig:Diagrams} [see how the two diagrams for $n=2$ correspond to the two leading-order terms in (\ref{eqn:DeltaS2Part})].
For each diagram contributing to the $n$th moment:
\begin{enumerate}
\item[1.]Assign $\mathcal{E}'^{i_{\ell+1}-i_\ell-1}\mathcal{Q}_*$ to each solid line.
\item[2.]Insert $\mathcal{P}_*=|\rho_*)(1|$ for each space between disconnected points.
\item[3.]Turn each (group of) dot(s) ``$\bullet$'' (connected by dashed lines) into $\tilde{\mathcal{E}}^{(m)}$ (where $m$ is the number of connected dots) while ``$\circ$'' into $|\rho_0)$.
\item[4.]Close each diagram with a trace $(1|$ at the left end.
\item[5.]Put $\frac{1}{N^n}\sum\cdots\sum$ to sum the contributions over all possible distances between nondegenerate points respecting the chronological ordering of the points, with an appropriate coefficient counting how many times such a diagram (the specific ordering of the points) appears in the original full range summation $\frac{1}{N^n}\sum\cdots\sum$ exploring all possible orderings of the points.
The right coefficient reads $n!/m_1!\,m_2!\cdots$, where $m_i$ are the numbers of coincident points connected by dashed lines in the relevant diagram and the factors $m_i!$ are to disregard the orderings among the coincident points. 
\end{enumerate}

It is easily recognized from Fig.\ \ref{fig:Diagrams} that the maximum number of $\mathcal{P}_*$ we can insert for the $n$th moment is given by $\lfloor\frac{n}{2}\rfloor$ (where now $\lfloor x\rfloor$ denotes the largest integer not greater than $x$).
Therefore, the substitution rules in (\ref{eqn:ReplaceRule}) tell us that the $n$th moment scales as anticipated in (\ref{eqn:Scaling}) (the power of $N$ is obtained by $n-\lfloor\frac{n}{2}\rfloor=\lceil\frac{n}{2}\rceil$).
As discussed in the beginning of this subsection, this is the right scaling for the central limit theorem, and only the even moments ($n=2,4,6,\ldots$) are relevant.
An important observation is that the leading-order contributions to the even moments are independent of the initial state $\rho_0$, since ``$\circ$'' representing the initial state $|\rho_0)$ is always disconnected from the first ``$\bullet$''.
See Fig.\ \ref{fig:Diagrams}.

\begin{figure}
\includegraphics[width=0.45\textwidth]{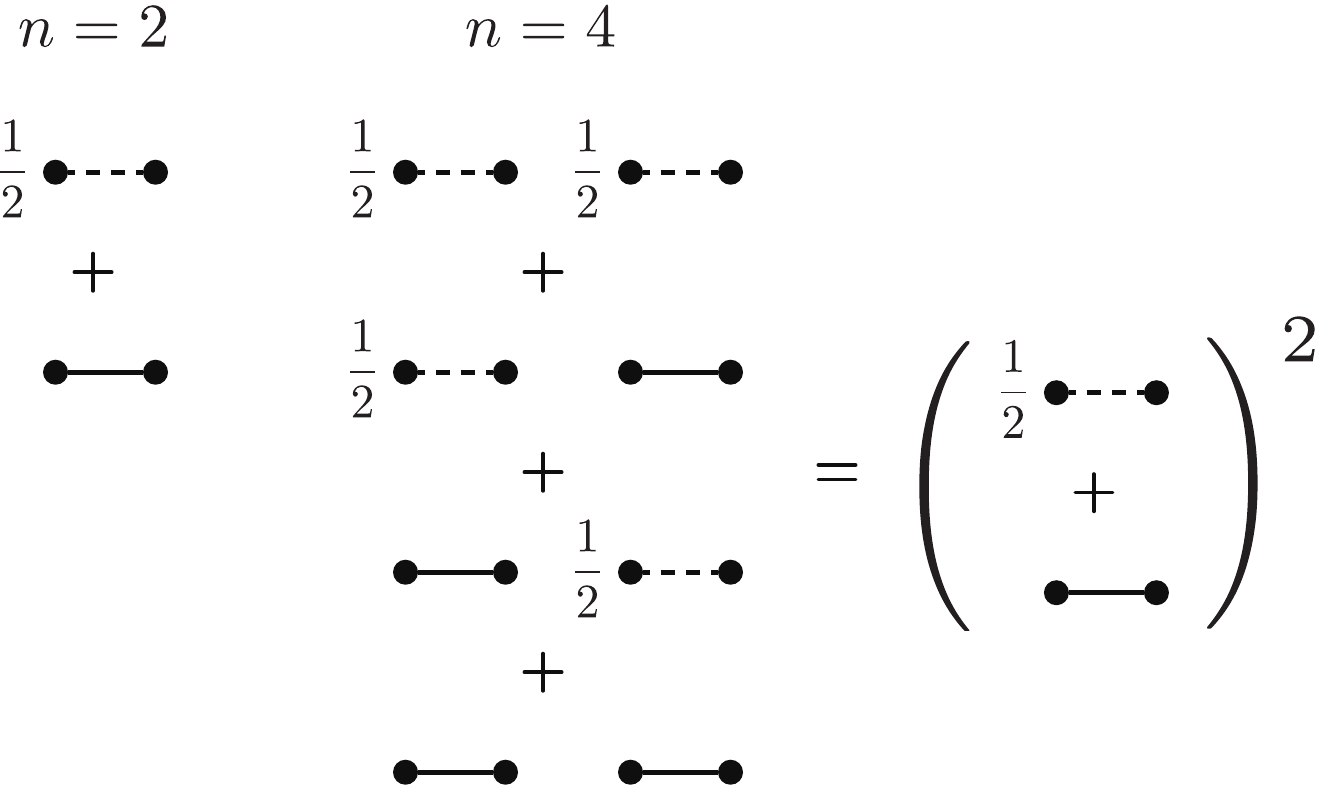}
\caption{The leading-order contributions to the even moments factorize.}
\label{fig:DiagramsFactorized}
\end{figure}

Let us look more carefully at the fourth moment.
The leading-order contributions represented by the diagrams in Fig.\ \ref{fig:Diagrams} read
\begin{align}
&\langle(S-\langle S\rangle_*)^4\rangle_N
\nonumber\\
&=\frac{4!}{(2!)^2}
\frac{1}{N^4}
\sum_{i_3=2}^N\sum_{i_1=1}^{i_3-1}
(1|
\tilde{\mathcal{E}}^{(2)}
|\rho_*)(1|
\tilde{\mathcal{E}}^{(2)}
|\rho_*)
\nonumber\displaybreak[0]\\
&\quad
{}+\frac{4!}{2!}
\frac{1}{N^4}
\sum_{i_3=3}^N\sum_{i_2=2}^{i_3-1}\sum_{i_1=1}^{i_2-1}
(1|
\tilde{\mathcal{E}}^{(2)}
|\rho_*)
\nonumber\\[-2.5truemm]
&\qquad\qquad\qquad\qquad\qquad\ %
{}\times
(1|
\tilde{\mathcal{E}}^{(1)}
\mathcal{E}'^{i_2-i_1-1}\mathcal{Q}_*
\tilde{\mathcal{E}}^{(1)}
|\rho_*)
\nonumber\displaybreak[0]\\
&\quad
{}+\frac{4!}{2!}
\frac{1}{N^4}
\sum_{i_4=3}^N\sum_{i_3=2}^{i_4-1}\sum_{i_1=1}^{i_3-1}
(1|
\tilde{\mathcal{E}}^{(1)}
\mathcal{E}'^{i_4-i_3-1}
\mathcal{Q}_*
\tilde{\mathcal{E}}^{(1)}
|\rho_*)
\nonumber\\[-2.5truemm]
&\qquad\qquad\qquad\qquad\qquad\qquad\qquad\qquad\qquad
{}\times
(1|
\tilde{\mathcal{E}}^{(2)}
|\rho_*)
\nonumber\displaybreak[0]\\
&\quad
{}+4!
\frac{1}{N^4}
\sum_{i_4=4}^N\sum_{i_3=3}^{i_4-1}\sum_{i_2=2}^{i_3-1}\sum_{i_1=1}^{i_2-1}
(1|
\tilde{\mathcal{E}}^{(1)}
\mathcal{E}'^{i_4-i_3-1}
\mathcal{Q}_*
\tilde{\mathcal{E}}^{(1)}
|\rho_*)
\nonumber\\[-2.5truemm]
&\qquad\qquad\qquad\qquad\qquad\qquad
{}\times
(1|
\tilde{\mathcal{E}}^{(1)}
\mathcal{E}'^{i_2-i_1-1}
\mathcal{Q}_*
\tilde{\mathcal{E}}^{(1)}
|\rho_*)
\nonumber\displaybreak[0]\\
&\quad
{}+O(1/N^3)
\nonumber\displaybreak[0]\\
&
=\frac{4!}{(2!)^2}
\frac{1}{N^4}
\frac{1}{2}N(N-1)
(1|
\tilde{\mathcal{E}}^{(2)}
|\rho_*)(1|
\tilde{\mathcal{E}}^{(2)}
|\rho_*)
\nonumber\displaybreak[0]\\
&\quad
{}+\frac{4!}{2!}
\frac{1}{N^4}
\frac{1}{2}N(N-1)
(1|
\tilde{\mathcal{E}}^{(2)}
|\rho_*)
(1|
\tilde{\mathcal{E}}^{(1)}
\frac{Q_*}{1-\mathcal{E}'}
\tilde{\mathcal{E}}^{(1)}
|\rho_*)
\nonumber\displaybreak[0]\\
&\quad
{}+\frac{4!}{2!}
\frac{1}{N^4}
\frac{1}{2}
N(N-1)
(1|
\tilde{\mathcal{E}}^{(1)}
\frac{Q_*}{1-\mathcal{E}'}
\tilde{\mathcal{E}}^{(1)}
|\rho_*)(1|
\tilde{\mathcal{E}}^{(2)}
|\rho_*)
\nonumber\displaybreak[0]\\
&\quad
{}+4!
\frac{1}{N^4}
\frac{1}{2}N(N-1)
(1|
\tilde{\mathcal{E}}^{(1)}
\frac{Q_*}{1-\mathcal{E}'}
\tilde{\mathcal{E}}^{(1)}
|\rho_*)
\nonumber\\
&\qquad\qquad\qquad\qquad\qquad\quad
{}\times
(1|
\tilde{\mathcal{E}}^{(1)}
\frac{Q_*}{1-\mathcal{E}'}
\tilde{\mathcal{E}}^{(1)}
|\rho_*)
\nonumber\displaybreak[0]\\
&\quad
{}+O(1/N^3).
\label{eqn:4thMoment}
\end{align}
This suggests that by the summations each of the leading-order contributions to an even moment $\langle(S-\langle S\rangle_*)^n\rangle_N$ acquires a common factor (binomial coefficient) $N!/(n/2)!(N-n/2)!$ while each $\mathcal{E}'^{i_{\ell+1}-i_\ell-1}$ is transformed into $(1-\mathcal{E}')^{-1}$.
Indeed, each leading-order diagram for an even moment consists of pairs of points $(i_{2r-1},i_{2r})$ ($r=1,\ldots,n/2$) connected by dashed or solid lines (see Fig.\ \ref{fig:Diagrams}), and its evaluation proceeds two points by two points with the help of the following formulas for $r=1,\ldots,n/2$ (with a convention $i_{n+1}=N+1$): for pairs of coincident points connected by dashed lines
\begin{equation}
\sum_{i_{2r-1}=r}^{i_{2r+1}-1}\frac{(i_{2r-1}-1)!}{(r-1)!(i_{2r-1}-r)!}
=\frac{(i_{2r+1}-1)!}{r!(i_{2r+1}-r-1)!}
\label{eqn:SumFormula1}
\end{equation}
[we have a single sum for each coincident pair: see (\ref{eqn:4thMoment})], while for pairs of nondegenerate points connected by solid lines
\begin{align}
&\sum_{i_{2r}=r+1}^{i_{2r+1}-1}
\sum_{i_{2r-1}=r}^{i_{2r}-1}
\mathcal{E}'^{i_{2r}-i_{2r-1}-1}
\frac{(i_{2r-1}-1)!}{(r-1)!(i_{2r-1}-r)!}
\nonumber\displaybreak[0]\\
&\quad
=\sum_{i_{2r-1}=r}^{i_{2r+1}-1}
\frac{1-\mathcal{E}'^{i_{2r+1}-i_{2r-1}-1}}{1-\mathcal{E}'}
\frac{(i_{2r-1}-1)!}{(r-1)!(i_{2r-1}-r)!}
\nonumber\displaybreak[0]\\
&\quad
=\frac{(i_{2r+1}-1)!}{r!(i_{2r+1}-r-1)!}
\frac{1}{1-\mathcal{E}'}
+O(i_{2r+1}^{r-1}).
\label{eqn:SumFormula2}
\end{align}
[The actual ranges of the summations in the leading-order contributions to the even moments are slightly different from those in (\ref{eqn:SumFormula1}) and (\ref{eqn:SumFormula2}), but the corrections are finite and become negligible in the asymptotic regime $N\gg n$.]
In this way, each leading-order diagram for an even moment acquires the binomial coefficient $N!/(n/2)!(N-n/2)!$ with $\mathcal{E}'^{i_{\ell+1}-i_\ell-1}$ being transformed into $(1-\mathcal{E}')^{-1}$.
This leads us to the following recipe for obtaining the expressions for the leading-order contributions to any even moment directly from the relevant diagrams: 
\begin{enumerate}
\item[$1'$.]Assign $(1|\tilde{\mathcal{E}}^{(2)}|\rho_*)/2$ to each pair of points connected by a dashed line.
\item[$2'$.]Assign $(1|\tilde{\mathcal{E}}^{(1)}(1-\mathcal{E}')^{-1}\mathcal{Q}_*\tilde{\mathcal{E}}^{(1)}|\rho_*)$ to each pair of points connected by a solid line.
\item[$3'$.]Give a common factor $n!N!/N^n(n/2)!(N-n/2)!\sim n!/N^{n/2}(n/2)!$ to each diagram.
\end{enumerate}
Then, the leading-order contributions to the even moments factorize as shown in Fig.\ \ref{fig:DiagramsFactorized} and yield (\ref{eqn:EvenMoments}).

\section{Use of Correlations}
\label{sec:Correlation}
An important difference from the standard strategy for parameter estimation, where independent identical experiments are performed to collect data, is that in the present sequential scheme the correlations among the measurement data are available for estimation.
Combining the information attainable from the correlations with that from the average $S$, the precision of the estimation can be enhanced.
The primary motivation of the present paper is to explore this possibility.

For instance, one can compute 
\begin{equation}
C_\ell
=\frac{1}{N-\ell}\sum_{i=1}^{N-\ell}s_is_{i+\ell}
\label{eqn:C}
\end{equation}
from a single sequence of $N$ measurement outcomes $\{s_1,\ldots,s_N\}$, which captures the correlation between two data separated by a distance $\ell$.
In the presence of the correlations among the data, $C_\ell$ may depend on the target parameter $g$ in a way that cannot be deduced solely from $S$.
This might provide additional knowledge on how the parameter $g$ is encoded in the process and can enhance the precision of the estimation of $g$.

In principle, $\ell$ ranges $\ell=1,\ldots,N-1$, but recall that the correlation between two data  are expected to decay exponentially as $\ell$ increases under a mixing channel: $C_\ell$ with $\ell$ greater than the correlation length would not contain useful information.
In addition, $N$ should be much greater than $\ell$ so that the number $N-\ell$ of data used to evaluate $C_\ell$ is large enough.
Therefore, we will require $N\gg L\ge\ell$,  with $L$ being the maximum $\ell$ we take to estimate the parameter $g$.

\begin{figure}[t]
\includegraphics[width=0.45\textwidth]{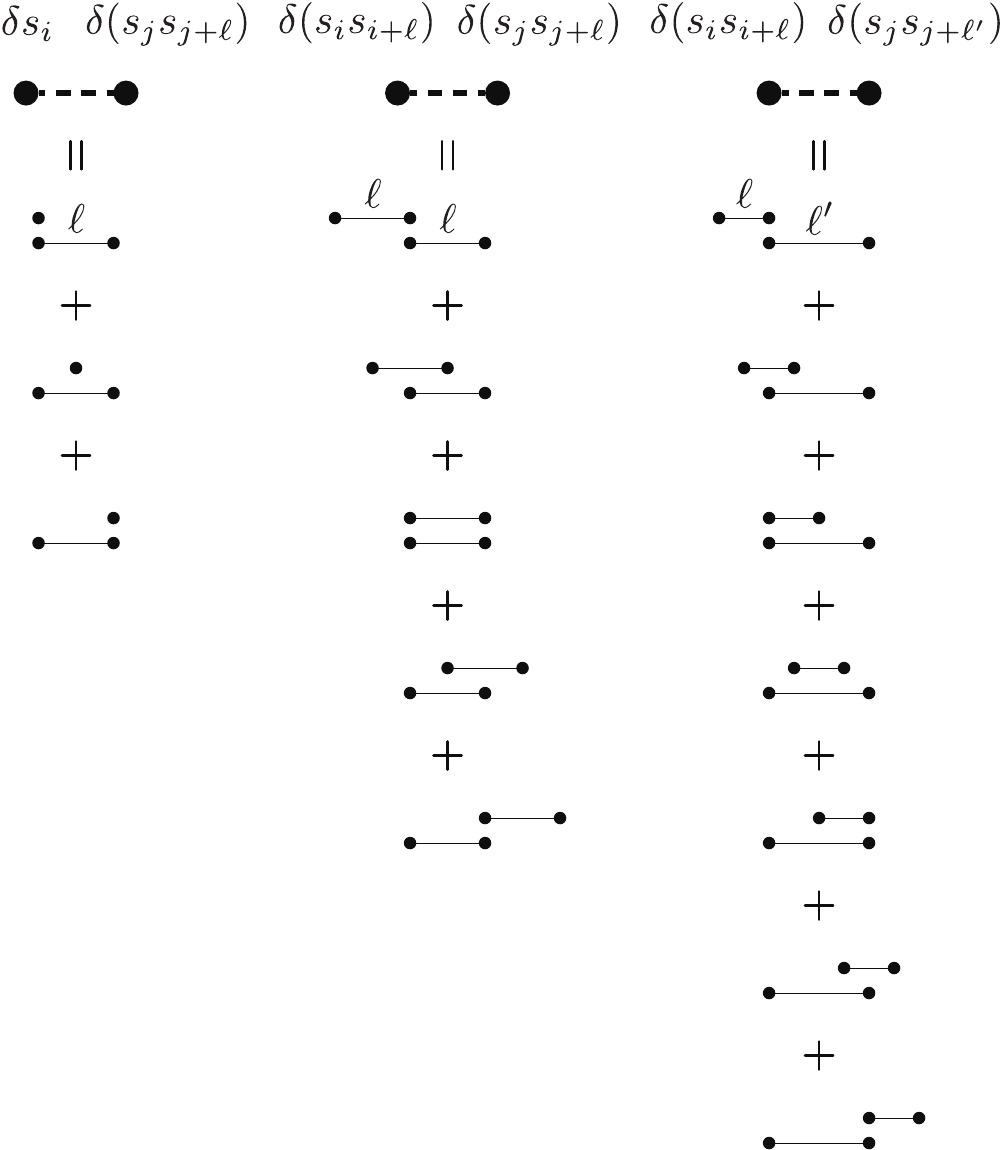}
\caption{Each pair of points $\delta(s_is_{i+\ell})_\ell$ is also represented by a dot ``$\bullet$'', and if the pair overlaps with another pair or a point $\delta s_i$, we connect the couple of dots ``$\bullet$'' by a dashed line.
Several terms are involved in such a single connected diagram as shown here.
These diagrams represent $\tilde{\mathcal{E}}_\ell^{(2)}$ and $\tilde{\mathcal{E}}_{\ell\ell'}^{(2)}$ in (\ref{eqn:E2l}) and (\ref{eqn:E2ll}).}
\label{fig:E2Corr}
\end{figure}

The correlations $C_\ell$ are also self-averaging quantities.
Moreover, we are able to prove that the central limit theorem holds for the set of quantities $\bm{X}=(S,C_1,\ldots,C_L)$.
First, the expectation value of $C_\ell$ is evaluated as
\begin{align}
\langle C_\ell\rangle_N
&=\sum_{s_1}\cdots\sum_{s_N}C_\ell\,p(s_1,\ldots,s_N|\rho_0)
\nonumber\displaybreak[0]\\
&
=\frac{1}{N-\ell}\sum_{i=1}^{N-\ell}\sum_{s_1}\cdots\sum_{s_N}s_is_{i+\ell}(1|\mathcal{E}_{s_N}\cdots\mathcal{E}_{s_1}|\rho_0)
\nonumber\displaybreak[0]\\
&
=\frac{1}{N-\ell}\sum_{i=1}^{N-\ell}
(1|\mathcal{E}^{(1)}\mathcal{E}^{\ell-1}\mathcal{E}^{(1)}\mathcal{E}^{i-1}|\rho_0)
\nonumber\displaybreak[0]\\
&=\langle C_\ell\rangle_*
+\frac{1}{N-\ell}
(1|\mathcal{E}^{(1)}\mathcal{E}^{\ell-1}\mathcal{E}^{(1)}\frac{1-\mathcal{E}'^{N-\ell}}{1-\mathcal{E}'}
\mathcal{Q}_*|\rho_0)
\nonumber\displaybreak[0]\\
&\xrightarrow{N\to\infty}
\langle C_\ell\rangle_*,
\end{align}
which approaches 
\begin{equation}
\langle C_\ell\rangle_*
=(1|\mathcal{E}^{(1)}\mathcal{E}^{\ell-1}\mathcal{E}^{(1)}|\rho_*)
=\langle s_is_{i+\ell}\rangle_*
\end{equation}
under the ergodicity of the channel $\mathcal{E}$.
Then, let us look at the $n$th moment
\begin{equation}
\mu_n(\bm{k})
=\left\langle
\biggl(
k_0(S-\langle S\rangle_*)+\sum_{\ell=1}^Lk_\ell(C_\ell-\langle C_\ell\rangle_*)
\biggr)^n
\right\rangle_N.
\end{equation}
It is an $n$th-order polynomial of $\bm{k}=(k_0,\ldots,k_L)$, and is a collection of all the $n$th moments among $\bm{X}=(S,C_1,\ldots,C_L)$ as its coefficients.
Since we are interested in the asymptotic limit $N\to\infty$, we collect the leading-order contributions to $\mu_n(\bm{k})$ for large $N$.
The idea to do that is basically the same as that for $S$: we try to insert $\mathcal{P}_*$ as many as possible in place of $\mathcal{E}$ between points $\delta s_i$ from $S-\langle S\rangle_*$ and pairs of points
\begin{equation}
\delta(s_is_{i+\ell})_\ell
=s_is_{i+\ell}-\langle C_\ell\rangle_*
\end{equation}
from $C_\ell-\langle C_\ell\rangle_*$.
Since we have 
$\langle\delta s_i\rangle_*=(1|\tilde{\mathcal{E}}^{(1)}|\rho_*)=0$ [Eq.\ (\ref{eqn:VanishExp})] and
\begin{equation}
\langle
\delta(s_is_{i+\ell})_\ell
\rangle_*
=(1|\tilde{\mathcal{E}}_\ell^{(1)}|\rho_*)=0,
\end{equation}
where
\begin{equation}
\tilde{\mathcal{E}}_\ell^{(1)}
=\sum_s\sum_{s'}\delta(ss')_\ell\,\mathcal{E}_s\mathcal{E}^{\ell-1}\mathcal{E}_{s'},
\label{eqn:E1l}
\end{equation}
there should be at least two pieces [points $\delta s_i$ and/or pairs of points $\delta(s_is_{i+\ell})_\ell$]
between two $\mathcal{P}_*$.
We just have to generalize the diagrammatic rules in Fig.\ \ref{fig:SeqPoints}: we represent each pair of points $\delta(s_is_{i+\ell})_\ell$ by a dot ``$\bullet$'', too, and if the pair overlaps with another pair or a point $\delta s_i$ we connect the couple of dots ``$\bullet$'' by a dashed line (see Fig.\ \ref{fig:E2Corr}), while if it does not we leave it disconnected from or connect it with its adjacent dot by a solid line depending on whether $\mathcal{P}_*$ is inserted between them or not.
The ranges of the summations $\sum\cdots\sum$ exploring all possible distances between dots ``$\bullet$'' should be carefully arranged depending on whether the dots represent points $\delta s_i$ or pairs of points $\delta(s_is_{i+\ell})_\ell$, and some of the prefactors in $1/N^n$ are replaced by $1/(N-\ell)$, but such details become irrelevant in the asymptotic regime $N\gg n,L$.
Then, the analysis goes in the same way as before, the leading-order diagrams are again given by Fig.\ \ref{fig:Diagrams}, and the $n$th moment $\mu_n(\bm{k})$ for an even $n$ asymptotically factorizes pairwise as Fig.\ \ref{fig:DiagramsFactorized}, where the pair of dots ``$\bullet$'' connected by a dashed line or a solid line represents the collection of all pairwise combinations among $\delta s_i$ and $\delta(s_is_{i+\ell})_\ell$ ($\ell=1,\ldots,L$), with some care on the coefficients to distinguish different orderings of the pieces, i.e., give a coefficient $1/2$ to the pair connected by the solid line in Fig.\ \ref{fig:DiagramsFactorized} and collect contributions with different orderings of the pieces [see the second and third terms in $\Sigma_{00}$, $\Sigma_{0\ell}$, and $\Sigma_{\ell\ell'}$ in (\ref{eqn:Covs00})--(\ref{eqn:Covsll}) below].
We get
\begin{equation}
\mu_n(\bm{k})
=\begin{cases}
\smallskip
\displaystyle
\frac{n!}{(2N)^{n/2}(n/2)!}\left(
\sum_{\ell=0}^{L}
\sum_{\ell'=0}^{L}
k_\ell\Sigma_{\ell\ell'}k_{\ell'}
\right)^{n/2}\\
\smallskip
\qquad\qquad\qquad\quad\ \,
{}+O(1/N^{\frac{n}{2}+1})
\ \ \,
(n\ \text{even}),\\
O(1/N^{\lceil\frac{n}{2}\rceil})
\qquad\qquad\qquad\qquad\quad\ \ \,
(n\ \,\,\text{odd}),
\end{cases}
\label{eqn:MomentsAsymp}
\end{equation}
where
\begin{align}
\Sigma_{00}
&
=(1|\tilde{\mathcal{E}}^{(2)}|\rho_*)
+2(1|\tilde{\mathcal{E}}^{(1)}\frac{\mathcal{Q}_*}{1-\mathcal{E}'}\tilde{\mathcal{E}}^{(1)}|\rho_*)
=\sigma^2,
\label{eqn:Covs00}
\displaybreak[0]\\
\Sigma_{0\ell}
&
=\Sigma_{\ell0}
=(1|\tilde{\mathcal{E}}_{\ell}^{(2)}|\rho_*)
+(1|\tilde{\mathcal{E}}^{(1)}\frac{\mathcal{Q}_*}{1-\mathcal{E}'}\tilde{\mathcal{E}}_\ell^{(1)}|\rho_*)
\nonumber\\
&\qquad\qquad\qquad\qquad\ \ %
{}+(1|\tilde{\mathcal{E}}_\ell^{(1)}\frac{\mathcal{Q}_*}{1-\mathcal{E}'}\tilde{\mathcal{E}}^{(1)}|\rho_*),
\label{eqn:Covs0l}
\displaybreak[0]\\
\Sigma_{\ell\ell'}
&
=(1|\tilde{\mathcal{E}}_{\ell\ell'}^{(2)}|\rho_*)
+(1|\tilde{\mathcal{E}}_\ell^{(1)}\frac{\mathcal{Q}_*}{1-\mathcal{E}'}\tilde{\mathcal{E}}_{\ell'}^{(1)}|\rho_*)
\nonumber\\
\displaystyle
&\qquad\qquad\qquad\qquad\ \ %
{}+(1|\tilde{\mathcal{E}}_{\ell'}^{(1)}\frac{\mathcal{Q}_*}{1-\mathcal{E}'}\tilde{\mathcal{E}}_\ell^{(1)}|\rho_*),
\label{eqn:Covsll}
\end{align}
with $\tilde{\mathcal{E}}^{(m)}$ and $\tilde{\mathcal{E}}_\ell^{(1)}$ defined in (\ref{eqn:En}) and (\ref{eqn:E1l}), respectively,
\begin{align}
\tilde{\mathcal{E}}_\ell^{(2)}
&=\sum_{s_1}\sum_{s_2}
(
\delta s_2
+
\delta s_1
)
\delta(s_2s_1)_\ell\,
\mathcal{E}_{s_2}\mathcal{E}^{\ell-1}\mathcal{E}_{s_1}
\nonumber\\
&\ \ %
{}+\sum_{k=1}^{\ell-1}\sum_{s_1}\sum_{s_2}\sum_{s_3}
\delta s_2\delta(s_3s_1)_\ell\,\mathcal{E}_{s_3}\mathcal{E}^{k-1}\mathcal{E}_{s_2}\mathcal{E}^{\ell-k-1}\mathcal{E}_{s_1}
\label{eqn:E2l}
\end{align}
and
\begin{align}
\tilde{\mathcal{E}}_{\ell\ell'}^{(2)}
={}&
\tilde{\mathcal{E}}_{\ell\ell'}^{\circ\star\bullet}
+\sum_{k=1}^{\min(\ell,\ell')-1}\tilde{\mathcal{E}}_{\ell\ell',k}^{\circ\bullet\circ\bullet}
\nonumber\\
&{}+\delta_{\ell\ell'}\tilde{\mathcal{E}}_{\ell}^{\star\star}
+(1-\delta_{\ell\ell'})
(
\tilde{\mathcal{E}}_{\ell\ell'}^{\star\circ\bullet}
+\tilde{\mathcal{E}}_{\ell\ell'}^{\bullet\circ\star}
+\tilde{\mathcal{E}}_{\ell\ell'}^{\bullet\circ\circ\bullet}
)
\label{eqn:E2ll}
\end{align}
composed of
\begin{equation}
\tilde{\mathcal{E}}_{\ell}^{\star\star}
=\sum_{s_1}
\sum_{s_2}
[\delta(s_2s_1)_{\ell}]^2
\mathcal{E}_{s_2}\mathcal{E}^{\ell-1}\mathcal{E}_{s_1},
\label{eqn:E2llcomp1}
\end{equation}
\begin{widetext}
\begin{align}
\tilde{\mathcal{E}}_{\ell\ell'}^{\circ\star\bullet}
&=\sum_{s_1}
\sum_{s_2}
\sum_{s_3}
\delta(s_3s_2)_\ell\,\delta(s_2s_1)_{\ell'}
\mathcal{E}_{s_3}\mathcal{E}^{\ell-1}\mathcal{E}_{s_2}\mathcal{E}^{\ell'-1}\mathcal{E}_{s_1}
+(\ell\leftrightarrow\ell'),
\\
\tilde{\mathcal{E}}_{\ell\ell',k}^{\circ\bullet\circ\bullet}
&=\sum_{s_1}
\sum_{s_2}
\sum_{s_3}
\sum_{s_4}
\delta(s_4s_2)_\ell\,\delta(s_3s_1)_{\ell'}
\mathcal{E}_{s_4}\mathcal{E}^{\ell-k-1}\mathcal{E}_{s_3}\mathcal{E}^{k-1}\mathcal{E}_{s_2}\mathcal{E}^{\ell'-k-1}\mathcal{E}_{s_1}
+(\ell\leftrightarrow\ell'),
\\
\tilde{\mathcal{E}}_{\ell\ell'}^{\star\circ\bullet}
&=\sum_{s_1}
\sum_{s_2}
\sum_{s_3}
\delta(s_3s_2)_{\min(\ell,\ell')}\delta(s_3s_1)_{\max(\ell,\ell')}
\mathcal{E}_{s_3}\mathcal{E}^{\min(\ell,\ell')-1}\mathcal{E}_{s_2}\mathcal{E}^{|\ell-\ell'|-1}\mathcal{E}_{s_1},
\displaybreak[0]\\
\tilde{\mathcal{E}}_{\ell\ell'}^{\bullet\circ\star}
&=\sum_{s_1}
\sum_{s_2}
\sum_{s_3}
\delta(s_3s_1)_{\max(\ell,\ell')}\delta(s_2s_1)_{\min(\ell,\ell')}
\mathcal{E}_{s_3}\mathcal{E}^{|\ell-\ell'|-1}\mathcal{E}_{s_2}\mathcal{E}^{\min(\ell,\ell')-1}\mathcal{E}_{s_1},
\displaybreak[0]\\
\tilde{\mathcal{E}}_{\ell\ell'}^{\bullet\circ\circ\bullet}
&=\sum_{k=1}^{|\ell-\ell'|-1}
\sum_{s_1}
\sum_{s_2}
\sum_{s_3}
\sum_{s_4}
\delta(s_3s_2)_{\min(\ell,\ell')}\delta(s_4s_1)_{\max(\ell,\ell')}
\mathcal{E}_{s_4}\mathcal{E}^{k-1}\mathcal{E}_{s_3}\mathcal{E}^{\min(\ell,\ell')-1}\mathcal{E}_{s_2}\mathcal{E}^{|\ell-\ell'|-k-1}\mathcal{E}_{s_1},
\label{eqn:E2llcomp6}
\end{align}
\end{widetext}
corresponding to the diagrams in Fig.\ \ref{fig:E2Corr}.
We provide the complete expressions for the covariances among $S$ and $C_\ell$ valid for any (even small) $N$ in Appendix \ref{app:CompCov}, whose asymptotic forms coincide with the covariances (\ref{eqn:Covs00})--(\ref{eqn:Covsll}) divided by $N$.

This result shows that the set of scaled variables $\sqrt{N}(S-\langle S\rangle_*)$ and $\sqrt{N}(C_\ell-\langle C_\ell\rangle_*)$ asymptotically become normal in the limit $N\to\infty$.
The characteristic function reads
\begin{align}
\chi(\bm{k})
&=\langle e^{i\sqrt{N}\sum_{\ell=0}^{L}k_\ell(X_\ell-\langle X_\ell\rangle_*)}\rangle_N
\nonumber\displaybreak[0]\\
&=\sum_{n=0}^\infty\frac{1}{n!}(i\sqrt{N})^n\mu_n(\bm{k})
\nonumber\displaybreak[0]\\
&\to
\sum_{r=0}^\infty
\frac{1}{r!}
\left(
-\frac{1}{2}
\sum_{\ell=0}^{L}
\sum_{\ell'=0}^{L}
k_\ell\Sigma_{\ell\ell'}k_{\ell'}
\right)^r
=e^{-\frac{1}{2}\bm{k}^T\Sigma\bm{k}},
\end{align}
where $\Sigma$ is the $(L+1)\times(L+1)$ matrix with its matrix elements given by the covariances in (\ref{eqn:Covs00})--(\ref{eqn:Covsll}).
The central limit theorem holds, and the probability distribution $P(\bm{X})$ of $\bm{X}=(S,C_1,\ldots,C_L)$ becomes asymptotically Gaussian,
\begin{equation}
P(\bm{X})
\simeq
\frac{
e^{-\frac{1}{2}N(\bm{X}-\langle\bm{X}\rangle_*)^T\Sigma^{-1}(\bm{X}-\langle\bm{X}\rangle_*)}
}{\sqrt{(2\pi/N)^{L+1}\det\Sigma}},
\label{eqn:PX}
\end{equation}
peaked at $\bm{X}=\langle\bm{X}\rangle_*$ with a shrinking covariance $\Sigma/N$.
This ensures that the single-shot values $\bm{X}$ computed from a single sequence of measurement data well represent their expectation values $\langle\bm{X}\rangle_*$, through which we can estimate a parameter $g$.
The uncertainty $\delta g$ in the estimation of $g$ is given by (\ref{eqn:CR}) with the Fisher information
\begin{align}
\mathcal{F}_L(g)
&=\int d^{L+1}\bm{X}\,P(\bm{X})\left(\frac{\partial}{\partial g}\ln P(\bm{X})\right)^2
\nonumber\displaybreak[0]\\
&
\simeq N\frac{\partial\langle\bm{X}\rangle_*^T}{\partial g}
\Sigma^{-1}\frac{\partial\langle\bm{X}\rangle_*}{\partial g},
\label{eqn:FSC}
\end{align}
which increases linearly in $N$, and the uncertainty $\delta g$ diminishes as $\delta g\simeq1/\sqrt{N}$ [in (\ref{eqn:FSC}) we have omitted the contribution from $\partial\Sigma/\partial g$ to the Fisher information $\mathcal{F}_L(g)$ since it does not grow with $N$].
Moreover, this Fisher information $\mathcal{F}_L(g)$ for the estimation of $g$ through a set of quantities $\bm{X}=(S,C_1,\ldots,C_L)$ can be greater than the Fisher information $\mathcal{F}(g)=\mathcal{F}_0(g)$ given in (\ref{eqn:FS}) for the estimation of the same $g$ but solely through $S$.
The precision of the estimation can be enhanced by looking at the correlation data $C_\ell$ in addition to the average $S$.

Here we have considered the two-point correlations $C_\ell$ as well as the average $S$.
If we incorporate higher-order correlations with more points, the precision of the estimation can be further improved.
On the other hand, correlations with too many points would not be helpful, since the number of data used to evaluate such correlations is reduced, and some of the points involved in the correlations are separated beyond the correlation length of the mixing channel supplying no more information than lower-order correlations.
It is currently not clear to what extent we can improve the precision of the estimation by looking at higher-order correlations.

\section{Example: Estimation of the Temperature of a Reservoir}\label{sec:ex}
\begin{figure}[b]
\includegraphics[width=0.29\textwidth]{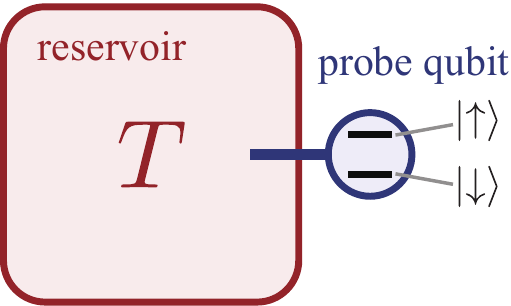}
\caption{We estimate the temperature of a thermal reservoir though measurements performed on a probe qubit in contact with the reservoir.}
\label{fig:QubitThermometer}
\end{figure}
In this section we analyze an explicit example, where the correlations among the data collected by the sequential measurements would be useful for improving the estimation of a parameter. 
The setting we consider is related to quantum thermometry, which aims to use low-dimensional quantum systems (say qubits) as temperature probes to minimize the undesired disturbance on the sample (see e.g.\ Refs.\ \cite{GERARDO,ref:LocalThermometerAnto} and references therein).
Specifically we focus on the paradigmatic example with a qubit probe in contact with a thermal reservoir at a finite temperature $T$. 
Our goal is to estimate the temperature $T$ of the reservoir by monitoring the relaxation dynamics induced on the qubit, which effectively plays the role of a local ``thermometer'' (Fig.\ \ref{fig:QubitThermometer}). 
In our approach we describe the probe-reservoir coupling in terms of the resulting  Markovian master equation \cite{ref:DynamicalMap-Alicki,ref:QuantumOptics-Scully,ref:OpenQuantumSystems,ref:QuantumNoise,ref:QuantumOptics-WallsMilburn} operating on the probe, i.e.,
\begin{align}
\frac{d}{dt}\rho(t)
={}&{-\frac{i}{2}}\Omega[\sigma_z,\rho(t)]
\nonumber\\
&{}-\frac{1}{2}\gamma_+[\sigma_+\sigma_-\rho(t)+\rho(t)\sigma_+\sigma_--2\sigma_-\rho(t)\sigma_+]
\nonumber\\
&{}-\frac{1}{2}\gamma_-[\sigma_-\sigma_+\rho(t)+\rho(t)\sigma_-\sigma_+-2\sigma_+\rho(t)\sigma_-],
\label{eqn:MasterEq}
\end{align}
where $\rho(t)$ represents the state of the qubit, $\hbar\Omega$ is the energy gap between the excited $\ket{\uparrow}$ and ground $\ket{\downarrow}$ states of the qubit, and
\begin{equation}
\sigma_z=\ket{\uparrow}\bra{\uparrow}-\ket{\downarrow}\bra{\downarrow},\quad
\sigma_+=\ket{\uparrow}\bra{\downarrow},\quad
\sigma_-=\ket{\downarrow}\bra{\uparrow}.
\end{equation}
The two relaxation constants $\gamma_+$ (for decay) and $\gamma_-$ (for excitation) are related to the temperature of the reservoir $T$, respecting the detailed balance condition.
For a bosonic thermal reservoir, they are given by \cite{ref:QuantumOptics-Scully,ref:OpenQuantumSystems,ref:QuantumNoise,ref:QuantumOptics-WallsMilburn}
\begin{equation}
\gamma_+=(1+n_\text{th})\gamma,\quad
\gamma_-=n_\text{th}\gamma,\quad
n_\text{th}=\frac{1}{e^{\hbar\Omega/k_BT}-1},
\end{equation}
with $k_B$ being the Boltzmann constant.
We assume that the parameters $\Omega$ and $\gamma$ (i.e., the characteristics of the thermometer) are known.
Estimating the temperature $T$ is then equivalent to estimating 
\begin{equation}
\gamma_\beta=\gamma_++\gamma_-
=\gamma\coth\frac{\hbar\Omega}{2k_BT},
\end{equation}
while $\gamma=\gamma_+-\gamma_-$ is a known constant independent of the temperature $T$.
The higher is the temperature, the larger is the decay rate $\gamma_\beta$.

\subsection{Standard Strategy}
The information about the temperature $T$, namely, the parameter $\gamma_\beta$, is imprinted in the state of the qubit through the dynamics under the influence of the thermal reservoir, i.e., by the action of the quantum channel $\Lambda_t$ which is the solution to the master equation (\ref{eqn:MasterEq}).
Then, the standard strategy to estimate the parameter $\gamma_\beta$ is 
\begin{enumerate}
\item
to prepare the qubit in a specific initial state $\rho_0$, 
\item to let the qubit evolve $\rho(\tau)=\Lambda_\tau(\rho_0)$ for a certain time $\tau$ in contact with the thermal reservoir, and
\item to measure a specific observable in the state $\rho(\tau)$.
\end{enumerate}
We repeat this experiment $N$ times to collect measurement results, from which we estimate the parameter $\gamma_\beta$.

\begin{figure*}
\begin{tabular}{l@{\qquad}l@{\qquad}l}
\includegraphics[width=0.3\textwidth]{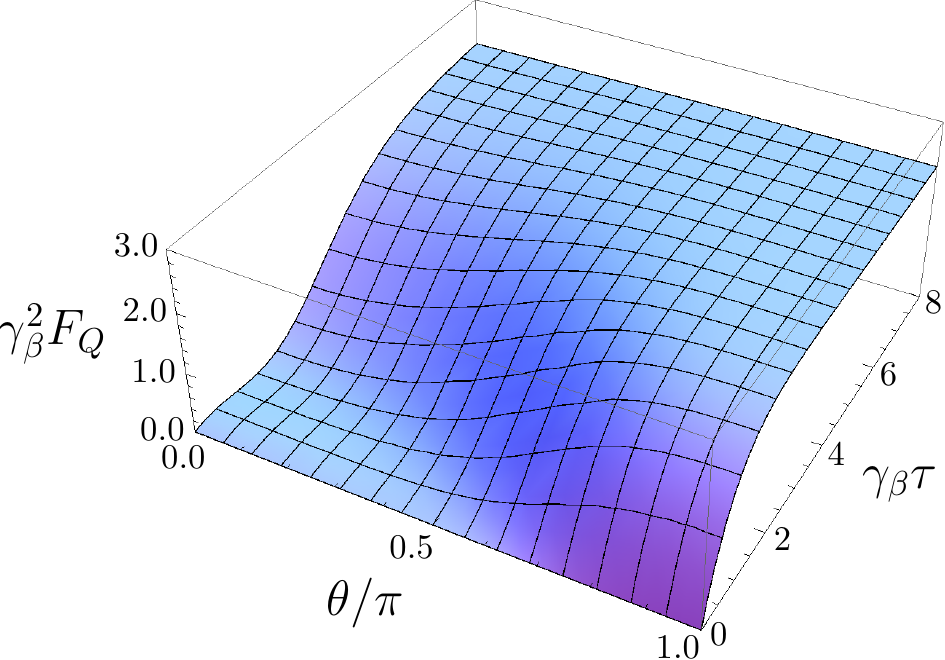}&
\includegraphics[width=0.3\textwidth]{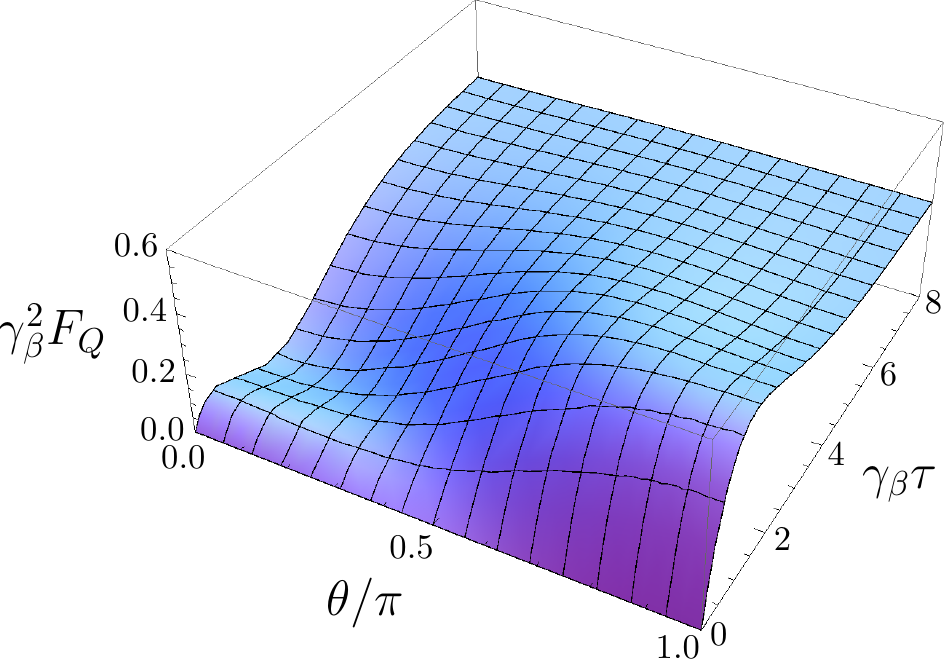}&
\includegraphics[width=0.3\textwidth]{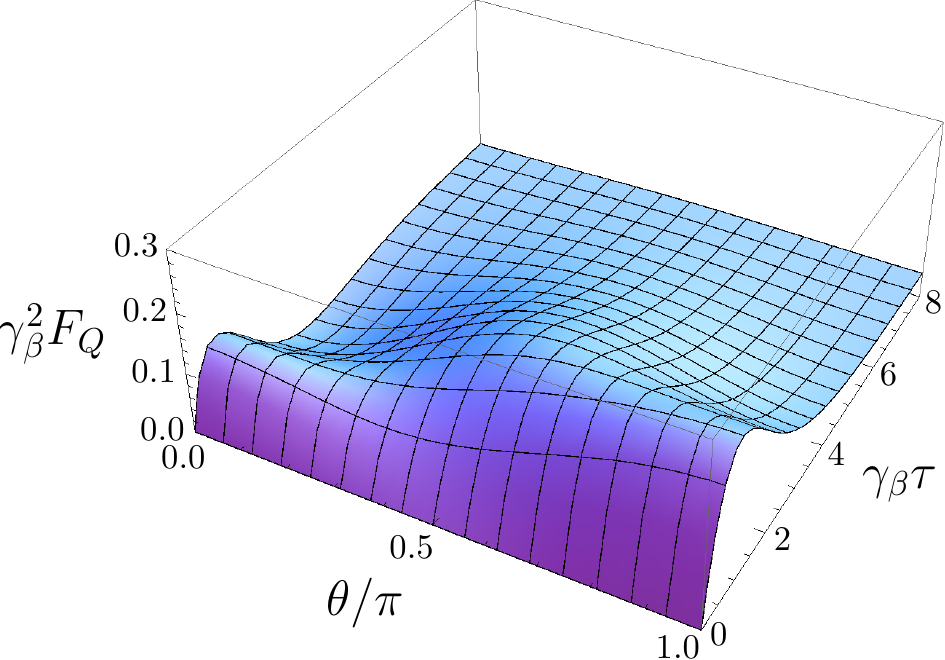}\\[-39truemm]
(a) $\gamma_\beta/\gamma=1.2$&
(b) $\gamma_\beta/\gamma=2.0$&
(c) $\gamma_\beta/\gamma=5.0$\\[35.5truemm]
\end{tabular}
\caption{The dependence of the quantum Fisher information $F_Q(\gamma_\beta)$ given in (\ref{eqn:Q0}) on the polar angle $\theta$ of a generic pure initial state $\ket{\psi_0}=e^{-i\varphi/2}\cos(\theta/2)\ket{\uparrow}+e^{i\varphi/2}\sin(\theta/2)\ket{\downarrow}$ and on the waiting time $\tau$, for different values of $\gamma_\beta/\gamma$, i.e., for different temperatures.
Note that $F_Q(\gamma_\beta)$ is symmetric around the polar axis and is independent of the azimuthal angle $\varphi$.}
\label{fig:Q0-ThetaTau}
\end{figure*}

For instance, we prepare the qubit in a specific initial state $\rho_0$, say in the excited state $\ket{\uparrow}$, and after a fixed waiting time $\tau$ we measure the qubit to check whether it is in the excited state $\ket{\uparrow}$ or in the ground state $\ket{\downarrow}$.
We repeat this process $N$ times, and we estimate $\gamma_\beta$ from the survival probability of the initial state $\ket{\uparrow}$ after time $\tau$.
Our measurement however can be weak and unsharp: here we consider the measurement which provokes the following back-action on the qubit,
\begin{equation}
\rho\mapsto\mathcal{M}_s(\rho)=M_s\rho M_s^\dag\quad
(s=\pm1)
\label{eqn:M}
\end{equation}
with
\begin{equation}
\begin{cases}
\medskip
M_{+1}=\cos\eta\,\ket{\uparrow}\bra{\uparrow}+\sin\eta\,\ket{\downarrow}\bra{\downarrow},\\
M_{-1}=\sin\eta\,\ket{\uparrow}\bra{\uparrow}+\cos\eta\,\ket{\downarrow}\bra{\downarrow},
\end{cases}
\label{eqn:Ms}
\end{equation}
depending on the outcome of the measurement $s$.
This measurement process can be simulated with a \textsc{cnot} gate \cite{ref:Wiseman-NJP-Uncertainty,ref:Steinberg-PRL-Uncertainty}.
The parameter $\eta$ controls the precision and the strength of the measurement: $\eta=0$ provides the projective measurement, while with $\eta=\pi/4$ the measurement gives totally random results with no disturbance on the measured system.
The probability of obtaining the measurement outcome $s$ in the state $\rho(\tau)$ is then given by
\begin{equation}
p_\tau(s|\rho_0)
=\Tr\{\mathcal{M}_s\bm{(}\rho(\tau)\bm{)}\}
=\Tr\{\Pi_s\rho(\tau)\},
\label{eqn:p_model}
\end{equation}
where
\begin{equation}
\Pi_s=M_s^\dag M_s=\begin{cases}
\medskip
\displaystyle
\cos^2\!\eta\,\ket{\uparrow}\bra{\uparrow}
+\sin^2\!\eta\,\ket{\downarrow}\bra{\downarrow}&
(s=+1),\\
\displaystyle
\sin^2\!\eta\,\ket{\uparrow}\bra{\uparrow}
+\cos^2\!\eta\,\ket{\downarrow}\bra{\downarrow}&
(s=-1)
\end{cases}
\label{eqn:POVM_model}
\end{equation}
are the POVM elements of this measurement.
The uncertainty in the estimation is then bounded by the Cram\'er-Rao inequality \cite{ref:Helstrom,ref:BraunsteinCave1994,ref:BraunsteinCave1996AnnPhys,ref:QuantumEstimation,ref:Paris-IJQI,ref:HolevoSNS,ref:MetrologyNaturePhoto,ref:Cramer}
\begin{equation}
\delta\gamma_\beta\gtrsim\frac{1}{\sqrt{NF(\gamma_\beta)}}
\end{equation}
with the Fisher information given by 
\begin{equation}
F(\gamma_\beta)
=\sum_{s=\pm1}p_\tau(s|\rho_0)
\left(
\frac{\partial}{\partial\gamma_\beta}\ln p_\tau(s|\rho_0)
\right)^2.
\label{eqn:F0}
\end{equation}

For the present model, the Bloch vector of the qubit evolves as
\begin{equation}
\begin{cases}
\medskip
\displaystyle
\langle\sigma_x\rangle_t
=e^{-\gamma_\beta t/2}\,
\Bigl(
\langle\sigma_x\rangle_0\cos\Omega t
-\langle\sigma_y\rangle_0\sin\Omega t
\Bigr),\\
\medskip
\displaystyle
\langle\sigma_y\rangle_t
=e^{-\gamma_\beta t/2}\,
\Bigl(
\langle\sigma_x\rangle_0\sin\Omega t
+\langle\sigma_y\rangle_0\cos\Omega t
\Bigr),\\
\displaystyle
\langle\sigma_z\rangle_t
=\left(
\langle\sigma_z\rangle_0
+\frac{\gamma}{\gamma_\beta}
\right)
e^{-\gamma_\beta t}
-\frac{\gamma}{\gamma_\beta},
\end{cases}
\end{equation}
where $\sigma_x=\sigma_++\sigma_-$ and $\sigma_y=-i(\sigma_+-\sigma_-)$.
The equilibrium state $\rho_\text{eq}$ is characterized by
\begin{equation}
\langle\sigma_x\rangle_\text{eq}
=\langle\sigma_y\rangle_\text{eq}
=0,\quad
\langle\sigma_z\rangle_\text{eq}
=-\frac{\gamma}{\gamma_\beta},
\end{equation}
namely,
\begin{equation}
\rho_\text{eq}
=\frac{1}{2}\left(
1-\frac{\gamma}{\gamma_\beta}\sigma_z
\right)
=\frac{
e^{-\beta\hbar\Omega\sigma_z/2}
}{
\Tr e^{-\beta\hbar\Omega\sigma_z/2}
}.
\label{eqn:eq}
\end{equation}
The probability distribution of the outcomes of the measurement (\ref{eqn:p_model}) at time $\tau$ reads
\begin{equation}
p_\tau(\pm1|\rho_0)
=\frac{1}{2}
\,\Bigl(
1\pm\langle\sigma_z\rangle_\tau\cos2\eta
\Bigr),
\end{equation}
and the Fisher information $F(\gamma_\beta)$ in (\ref{eqn:F0}) is estimated to be
\begin{equation}
F(\gamma_\beta)
=
\frac{
\cos^2\!2\eta
}{
1-\langle\sigma_z\rangle_\tau^2\cos^2\!2\eta
}
\left(
\frac{\partial\langle\sigma_z\rangle_\tau}{\partial\gamma_\beta}
\right)^2.
\label{eqn:F0_model}
\end{equation}

A larger Fisher information would be attainable by measuring a different observable.
The maximum Fisher information one can reach with the optimal measurement is given by the quantum Fisher information \cite{ref:QuantumMetrologyVittorio,ref:MetrologyNaturePhoto,ref:Helstrom,ref:BraunsteinCave1994,ref:BraunsteinCave1996AnnPhys,ref:QuantumEstimation,ref:Paris-IJQI,ref:HolevoSNS},
\begin{equation}
F_Q(\gamma_\beta)
=\Tr\{\rho(\tau)L_{\gamma_\beta}^2\}
=
\frac{\partial\langle\bm{\sigma}\rangle_\tau^T}{\partial\gamma_\beta}
V^{-1}
\frac{\partial\langle\bm{\sigma}\rangle_\tau}{\partial\gamma_\beta},
\label{eqn:Q0}
\end{equation}
with the symmetric logarithmic derivative 
\begin{equation}
L_{\gamma_\beta}
=
\frac{\partial\langle\bm{\sigma}\rangle_\tau^T}{\partial\gamma_\beta}
V^{-1}(\bm{\sigma}-\langle\bm{\sigma}\rangle_\tau),
\end{equation}
where $V$ is a $3\times3$ matrix whose matrix elements are given by
\begin{equation}
V_{ij}
=
\delta_{ij}
-\langle\sigma_i\rangle_\tau
\langle\sigma_j\rangle_\tau
\quad
(i,j=x,y,z).
\end{equation}

\begin{figure}
\includegraphics[width=0.4\textwidth]{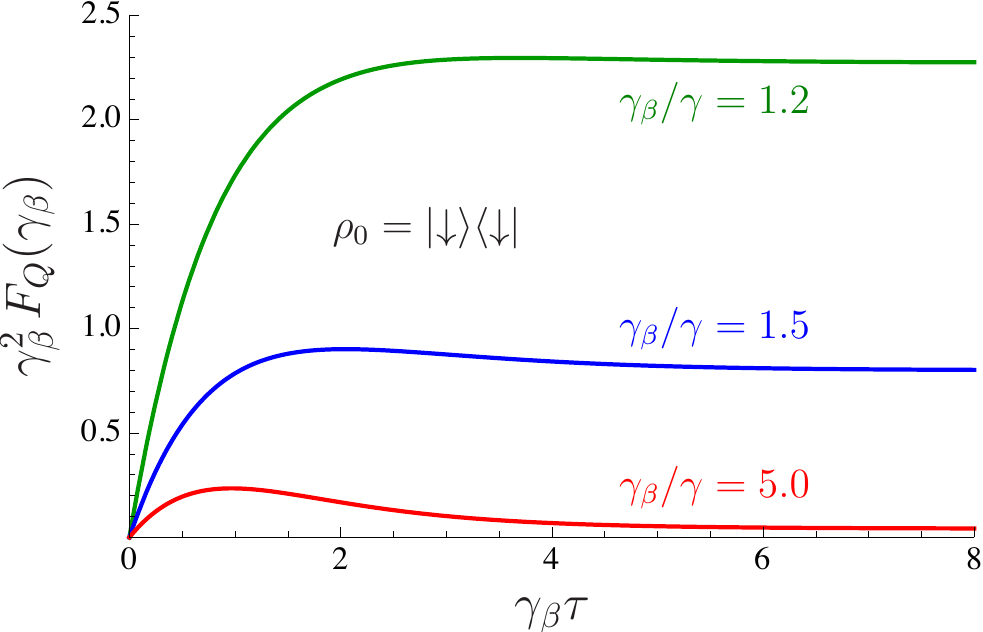}
\caption{The temporal behavior of the quantum Fisher information $F_Q(\gamma_\beta)$ given in (\ref{eqn:Q0}) for $\rho_0=\ket{\downarrow}\bra{\downarrow}$ and for different $\gamma_\beta$ (for different temperatures).}
\label{fig:Q0-ThetaPi}
\end{figure}

\begin{figure}
\includegraphics[width=0.4\textwidth]{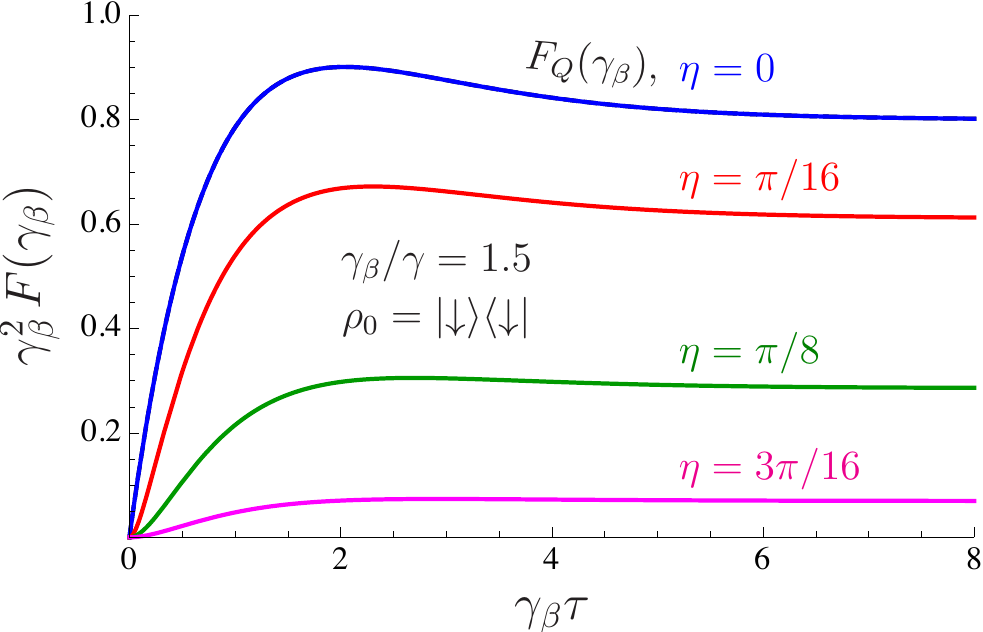}
\caption{The temporal behavior of the Fisher information $F(\gamma_\beta)$ given in (\ref{eqn:F0_model}) for $\rho_0=\ket{\downarrow}\bra{\downarrow}$ and $\gamma_\beta/\gamma=1.5$ with different strengths of the measurement $\eta$. In the case of projective measurement $\eta=0$, the Fisher information $F(\gamma_\beta)$ coincides with the quantum Fisher information $F_Q(\gamma_\beta)$ given in (\ref{eqn:Q0}) and plotted in Fig.\ \ref{fig:Q0-ThetaPi}.}
\label{fig:F0-TauEta}
\end{figure}

Notice here that both the Fisher information $F(\gamma_\beta)$ in (\ref{eqn:F0_model}) and the quantum Fisher information $F_Q(\gamma_\beta)$ in (\ref{eqn:Q0}) depend on the choice of the initial state $\rho_0$.
Because of the convexity of the quantum Fisher information, the maximum of the quantum Fisher information (the best estimation) is always achieved by choosing a pure input state $\rho_0=\ket{\psi_0}\bra{\psi_0}$ \cite{ref:QFI-convexity-Fujiwara2001,ref:MetrologyNaturePhoto}.
Moreover, for the present problem, the ground state of the qubit $\ket{\psi_0}=\ket{\downarrow}$ is the optimal choice, in the sense that the maximum of $F_Q(\gamma_\beta)$ for a given temperature is achieved with $\ket{\psi_0}=\ket{\downarrow}$: see Fig.\ \ref{fig:Q0-ThetaTau}, and the temporal behavior of $F_Q(\gamma_\beta)$ for $\ket{\psi_0}=\ket{\downarrow}$ is plotted in Fig.\ \ref{fig:Q0-ThetaPi}.
For this specific initial state $\rho_0=\ket{\downarrow}\bra{\downarrow}$, the Fisher information $F(\gamma_\beta)$ in (\ref{eqn:F0_model}) with $\eta=0$ coincides with the quantum Fisher information $F_Q(\gamma_\beta)$ in (\ref{eqn:Q0}), for any time $\tau$ and for any $\gamma_\beta$: the projective measurement to discriminate $\ket{\uparrow}$ and $\ket{\downarrow}$ is the optimal measurement.
For nonvanishing $\eta>0$ the Fisher information $F(\gamma_\beta)$ is reduced, and the weaker is the measurement, the smaller is the Fisher information $F(\gamma_\beta)$, as shown in Fig.\ \ref{fig:F0-TauEta}.

\subsection{Sequential Scheme}
Let us now turn our attention to the sequential scheme.
First, it is important to check whether the channel $\mathcal{E}$ defined in (\ref{eqn:E}) with $\mathcal{E}_s$ in (\ref{eqn:Evo}) is mixing.
For the present model, the spectrum of $\mathcal{E}$ is given by 
\begin{equation}
\{1,e^{-\gamma_\beta\tau},e^{-(\gamma_\beta/2\pm i\Omega)\tau}\}, 
\label{eqn:spectrum_model}
\end{equation}
and therefore, $\mathcal{E}$ is mixing for any $\tau>0$ with a unique fixed point (the eigenstate belonging to the eigenvalue $1$)
\begin{equation}
\rho_*=\rho_\text{eq},
\label{eqn:FixedPointThEq}
\end{equation}
which coincides with the equilibrium state $\rho_\text{eq}$ in (\ref{eqn:eq}) of the free relaxation process.
This mixing is apparently a direct consequence of the irreversibility of the relaxation process $\Lambda_t$ of the probe qubit.
Since $\mathcal{E}$ is mixing, the sequential scheme works for the present problem.

Let us take the average of the outcomes of a sequence of $N$ measurements, $S$ defined in (\ref{eqn:S}), as a quantity through which we estimate $\gamma_\beta$.
For the present model, its expectation value is computed to be
\begin{align}
\langle S\rangle_N
&=
-\left[
\frac{\gamma}{\gamma_\beta}
-\frac{1}{N}
\frac{1-e^{-N\gamma_\beta\tau}}{e^{\gamma_\beta\tau}-1}
\left(
\langle\sigma_z\rangle_0
+\frac{\gamma}{\gamma_\beta}
\right)
\right]
\cos2\eta
\nonumber\\
&\qquad\qquad\qquad\qquad\qquad\qquad\qquad\qquad\qquad
(N\ge1)
\end{align}
and the variance to be 
\begin{widetext}
\begin{align}
(\Delta S)_N^2
=\langle S^2\rangle_N-\langle S\rangle_N^2
={}&
\frac{1}{N}\left[
\sin^2\!2\eta
+
\left(
\frac{1+e^{-\gamma_\beta\tau}}{1-e^{-\gamma_\beta\tau}}
-\frac{2}{N}\frac{1-e^{-N\gamma_\beta\tau}}{(1-e^{-\gamma_\beta\tau})^2}
e^{-\gamma_\beta\tau}
\right)
\left(
1-\frac{\gamma^2}{\gamma_\beta^2}
\right)
\cos^2\!2\eta
\right]
\nonumber\\
&{}-\frac{1}{N}
\left(
\frac{
e^{-N\gamma_\beta\tau}
}{
1-e^{-N\gamma_\beta\tau}
}
-\frac{1}{2N}\frac{
1+e^{-\gamma_\beta\tau}
}{1-e^{-\gamma_\beta\tau}}
\right)
\frac{
1-e^{-N\gamma_\beta\tau}
}{e^{\gamma_\beta\tau}-1}
\left(
\langle\sigma_z\rangle_0
+\frac{\gamma}{\gamma_\beta}
\right)
\frac{\gamma}{\gamma_\beta}
\cos^2\!2\eta
\nonumber\displaybreak[0]\\
&{}-
\frac{1}{N^2}
\left[
\frac{1-e^{-N\gamma_\beta\tau}}{e^{\gamma_\beta\tau}-1}
\left(
\langle\sigma_z\rangle_0
+\frac{\gamma}{\gamma_\beta}
\right)
\cos2\eta
\right]^2
\qquad\qquad\qquad\qquad\qquad\quad
(N\ge1).
\end{align}
\end{widetext}
As $N$ increases, both become independent of the initial state $\rho_0$, and the variance $(\Delta S)_N^2$ shrinks as $1/N$,
\begin{gather}
\langle S\rangle_N\to-\frac{\gamma}{\gamma_\beta}\cos2\eta,
\label{eqn:SbarModel}
\displaybreak[0]\\
(\Delta S)_N^2
\to\frac{1}{N}\left[
\sin^2\!2\eta
+
\frac{1+e^{-\gamma_\beta\tau}}{1-e^{-\gamma_\beta\tau}}
\left(
1-\frac{\gamma^2}{\gamma_\beta^2}
\right)
\cos^2\!2\eta
\right].
\label{eqn:VarSModel}
\end{gather}
In other words, $S$ evaluated from a single sequence of measurements almost certainly exhibits a value very close to its expectation value $\langle S\rangle_N$, which is a function of $\gamma_\beta$.
Therefore, by comparing $S$ (obtained via a single experimental run) with its expectation value $\langle S\rangle_N$ [given by the formula (\ref{eqn:SbarModel})], the parameter $\gamma_\beta$ is estimated with the uncertainty regulated by the variance $(\Delta S)_N^2$ in (\ref{eqn:VarSModel}), i.e., with the precision given by the Fisher information $\mathcal{F}(\gamma_\beta)=\mathcal{F}_0(\gamma_\beta)$ in (\ref{eqn:FS}),
\begin{equation}
\mathcal{F}_0(\gamma_\beta)
\to
\frac{N}{\gamma_\beta^2}
\frac{\displaystyle
\frac{\gamma^2}{\gamma_\beta^2}
\cos^2\!2\eta
}{\displaystyle
\sin^2\!2\eta
+
\frac{1+e^{-\gamma_\beta\tau}}{1-e^{-\gamma_\beta\tau}}
\left(
1-\frac{\gamma^2}{\gamma_\beta^2}
\right)
\cos^2\!2\eta
},
\end{equation}
which is to be compared with the Fisher information $NF(\gamma_\beta)$ with (\ref{eqn:F0_model}) by the standard strategy (see Fig.\ \ref{fig:Fcomp} below).

As stressed above, the correlations among the acquired data are also available for the estimation in the sequential scheme.
For instance, the two-point correlations $C_\ell$ defined in (\ref{eqn:C}) can be used to estimate $\gamma_\beta$.
Their expectation values (for a generic initial state $\rho_0$) are given by
\begin{widetext}
\begin{multline}
\langle C_\ell\rangle_N
=
\left[
\frac{\gamma^2}{\gamma_\beta^2}
+
e^{-\ell\gamma_\beta\tau}
\left(
1-\frac{\gamma^2}{\gamma_\beta^2}
\right)
-\frac{1}{N-\ell}
\frac{1-e^{-(N-\ell)\gamma_\beta\tau}}{e^{\gamma_\beta\tau}-1}
(1-e^{-\ell\gamma_\beta\tau})
\left(
\langle\sigma_z\rangle_0
+\frac{\gamma}{\gamma_\beta}
\right)
\frac{\gamma}{\gamma_\beta}
\right]\cos^2\!2\eta
\\
(\ell\ge0,\ N\ge\ell+1),
\end{multline}
and their covariances (in the stationary state $\rho_0=\rho_*$) by
\begin{align}
&\langle C_\ell C_{\ell'}\rangle_N
-\langle C_\ell\rangle_N\langle C_{\ell'}\rangle_N
\vphantom{\frac{1}{N}}
\nonumber\\
&\quad
=\delta_{\ell\ell'}
\frac{1}{N}
\sin^4\!2\eta
+
\frac{2}{N}
(\langle C_{\ell-\ell'}\rangle_*+\langle C_{\ell+\ell'}\rangle_*)
\sin^2\!2\eta
\nonumber\displaybreak[0]\\
&\qquad
{}-
\frac{2}{N}
\left[
\ell'e^{-(\ell+\ell')\gamma_\beta\tau}
-
\frac{1}{2}
\left(
\ell-\ell'
+\frac{
1+e^{-2 \gamma_\beta\tau}
}{
1-e^{-2 \gamma_\beta\tau}
}
\right)
(
e^{-(\ell-\ell')\gamma_\beta\tau}
-e^{-(\ell+\ell')\gamma_\beta\tau}
)
\right]
\left(
1-\frac{\gamma^2}{\gamma_\beta^2}
\right)^2
\cos^4\!2\eta
\nonumber\displaybreak[0]\\
&\qquad
{}-\frac{4}{N}
\left[
\ell'
(
e^{-\ell\gamma_\beta\tau}
+e^{-\ell'\gamma_\beta\tau}
)
-\frac{
1+e^{-\gamma_\beta\tau}
}{
1-e^{-\gamma_\beta\tau}
}
(
1-e^{-\ell'\gamma_\beta\tau}
)
\right]
\left(
1-\frac{\gamma^2}{\gamma_\beta^2}
\right)
\frac{\gamma^2}{\gamma_\beta^2}
\cos^4\!2\eta
+O(1/N^2)
\qquad
(\ell\ge\ell'\ge1,\ N\gg\ell),
\label{eqn:CovCModelAsymp}\\
&\langle S C_{\ell}\rangle_N
-\langle S\rangle_N\langle C_{\ell}\rangle_N
\vphantom{\frac{1}{N}}
\nonumber\\
&\quad
=\frac{2}{N}\langle S\rangle_*
\,\Biggl\{
\sin^2\!2\eta
-\left[
\ell
e^{-\ell\gamma_\beta\tau}
-\left(
\frac{
1+e^{-\gamma_\beta\tau}
}{1-e^{-\gamma_\beta\tau}}
-\frac{1}{N-\ell}
\frac{1-e^{-(N-\ell)\gamma_\beta\tau}}{(1-e^{-\gamma_\beta\tau})^2}
e^{-\gamma_\beta\tau}
\right)
(1-e^{-\ell\gamma_\beta\tau})
\right]
\left(
1-\frac{\gamma^2}{\gamma_\beta^2}
\right)
\cos^2\!2\eta
\Biggr\}
\nonumber\\
&\qquad\qquad\qquad\qquad\qquad\qquad\qquad\qquad\qquad\qquad\qquad\qquad\qquad\qquad\qquad\qquad\qquad\qquad\qquad\qquad\qquad
(\ell\ge1,\ N\ge\ell+1)
\end{align}
\end{widetext}
[the complete expression for the covariance $\langle C_\ell C_{\ell'}\rangle_N-\langle C_\ell\rangle_N\langle C_{\ell'}\rangle_N$ valid for any $\ell,\ell'\ge1$ and $N\ge\max(\ell,\ell')+1$ (but for $\rho_0=\rho_*$) is given in Appendix \ref{sec:covariances}].
All the covariances scale as $1/N$, and the Fisher information (\ref{eqn:FSC}) increases linearly in $N$.
This ensures that, by comparing the set of quantities $(S,C_1,\ldots,C_L)$ evaluated from a single sequence of measurement data with the set of their expectation values $(\langle S\rangle_N,\langle C_1\rangle_N,\ldots,\langle C_L\rangle_N)$, one can estimate $\gamma_\beta$ with the precision given by the Fisher information $\mathcal{F}_L(\gamma_\beta)$ computed by the formula (\ref{eqn:FSC}), which increases linearly in $N$.
It is reasonable to expect that the estimation with the multiple quantities $(S,C_1,\ldots,C_L)$ is better in precision than the estimation solely through the average $S$, namely, the Fisher information $\mathcal{F}_L(\gamma_\beta)$ ($L>0$) is larger than the Fisher information $\mathcal{F}_0(\gamma_\beta)$, and the more correlations are incorporated (the larger is the number $L$), the larger is the Fisher information $\mathcal{F}_L(\gamma_\beta)$.

Let us look at two different regimes.
\begin{figure*}
\begin{tabular}{c@{\quad\ \ }c@{\quad\ \ }c}
(a)&(c)&(d)\\
\includegraphics[scale=0.53]{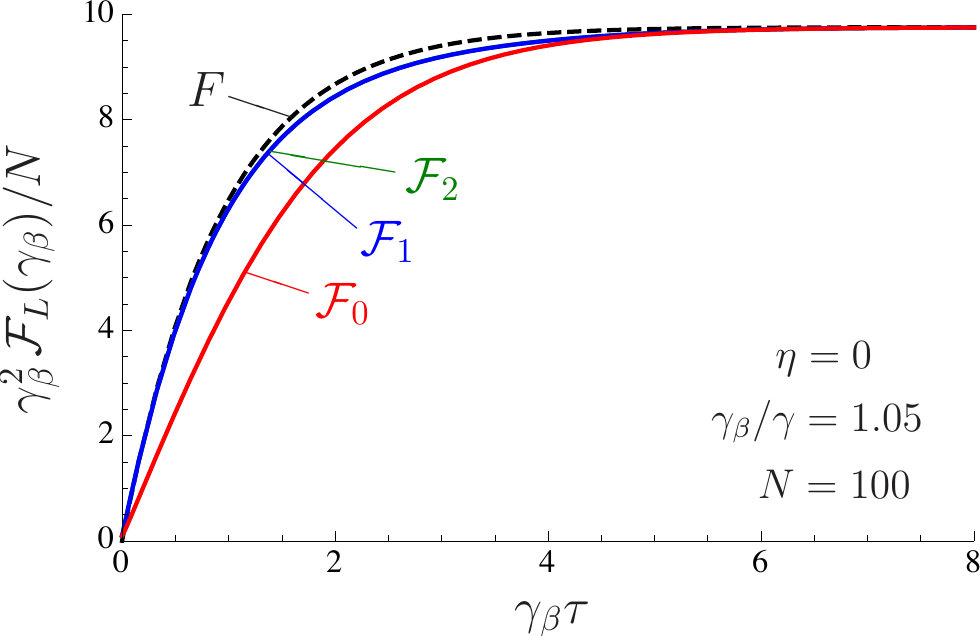}&
\includegraphics[scale=0.53]{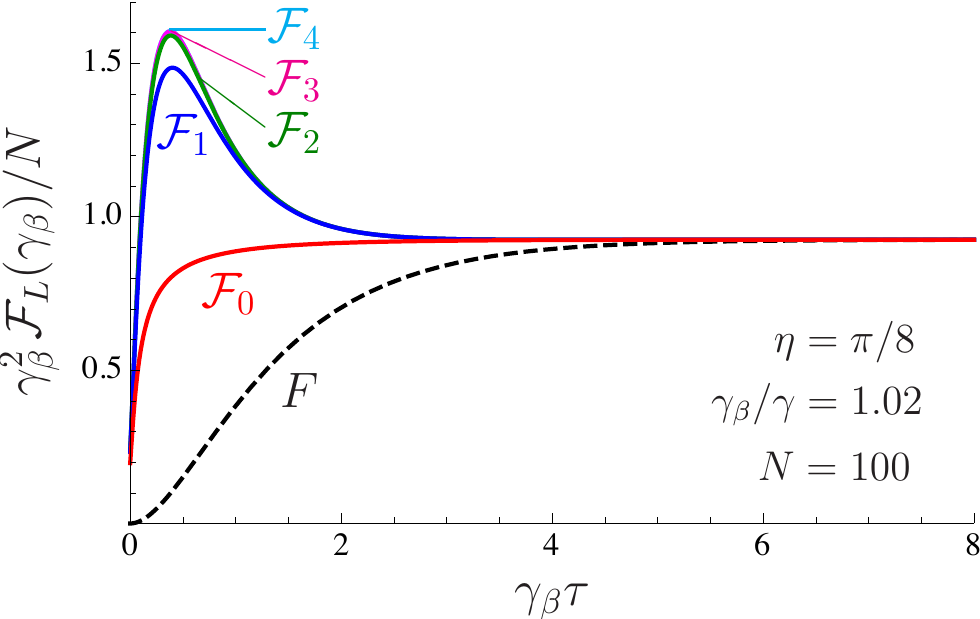}&
\includegraphics[scale=0.53]{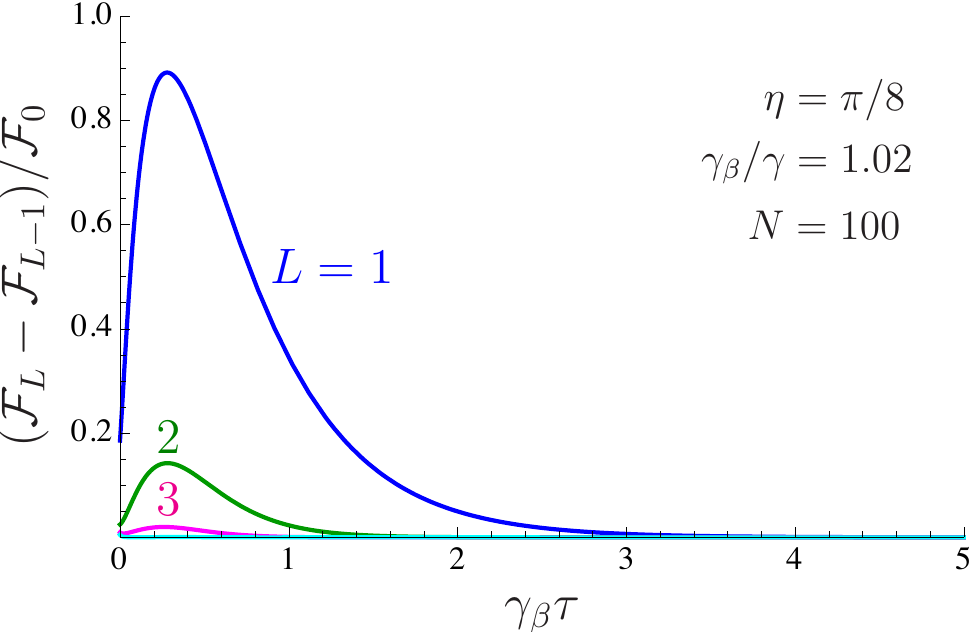}\\
\includegraphics[scale=0.53]{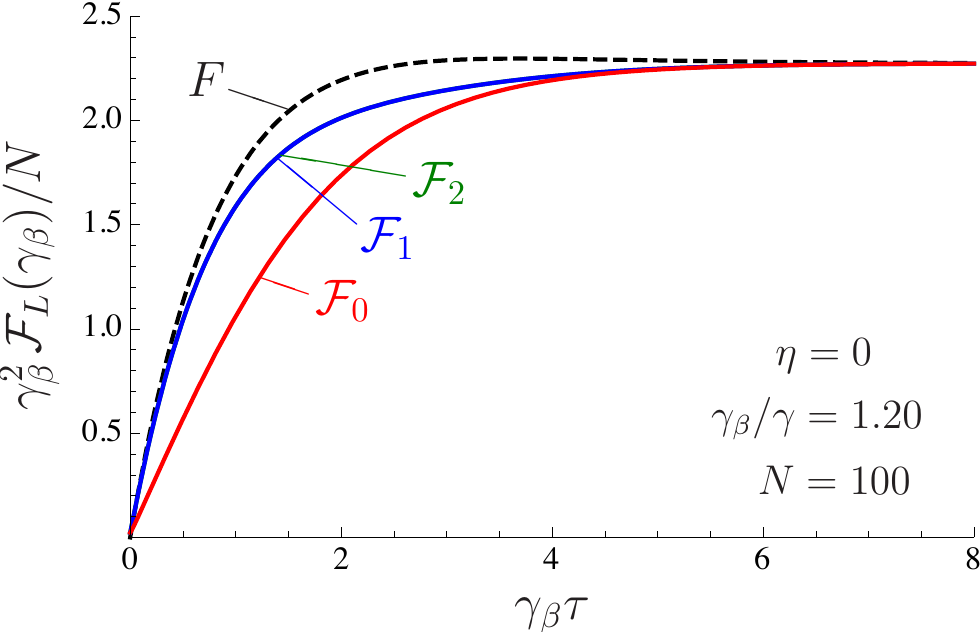}&
\includegraphics[scale=0.53]{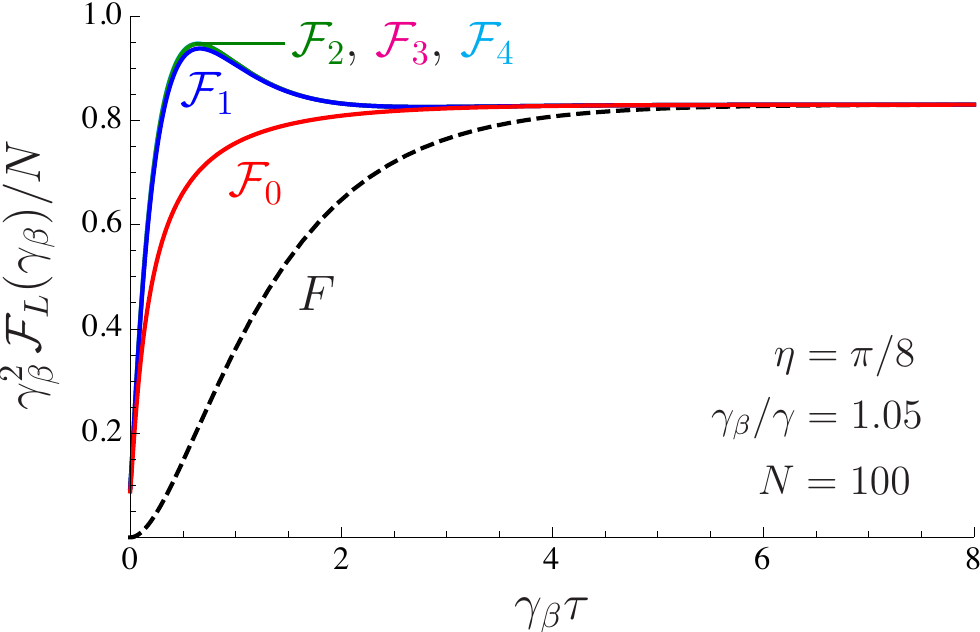}&
\includegraphics[scale=0.53]{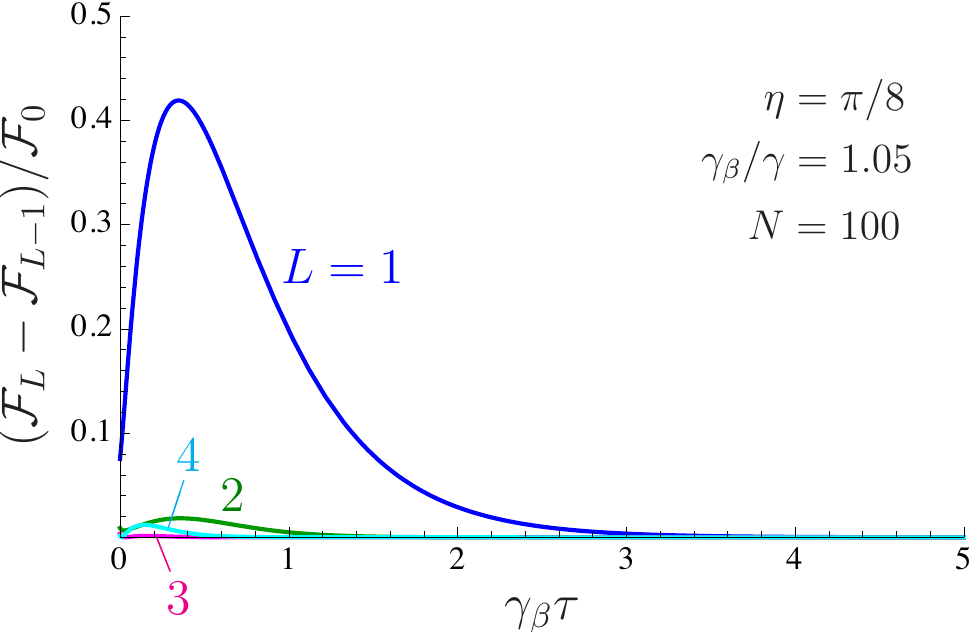}\\
\includegraphics[scale=0.53]{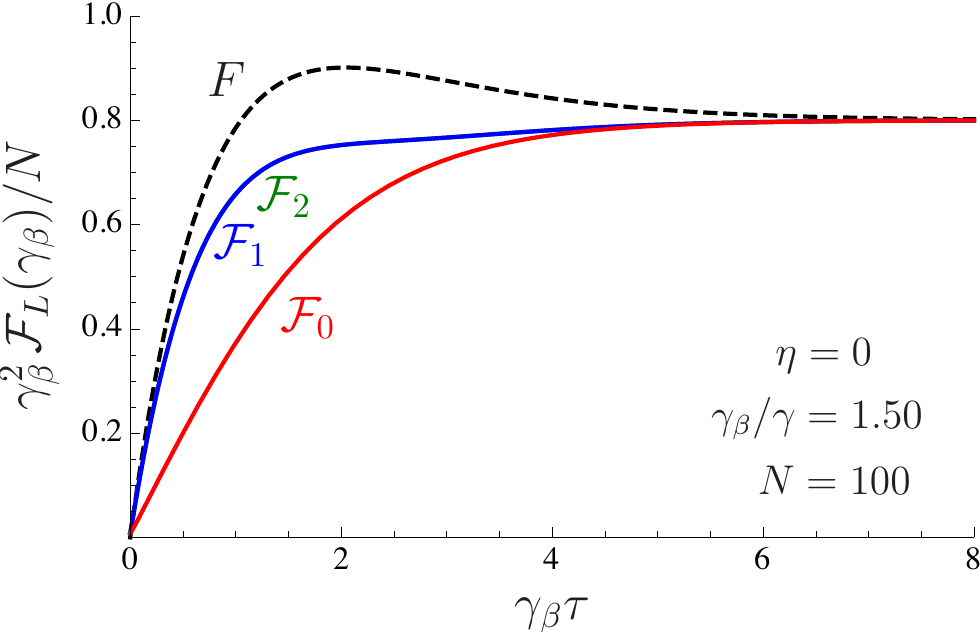}&
\includegraphics[scale=0.53]{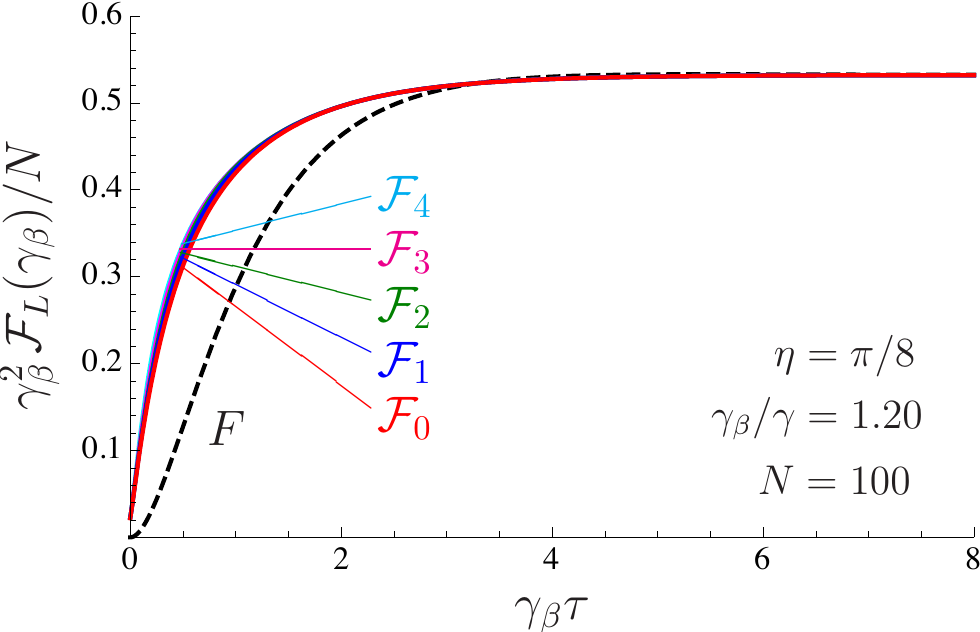}&
\includegraphics[scale=0.53]{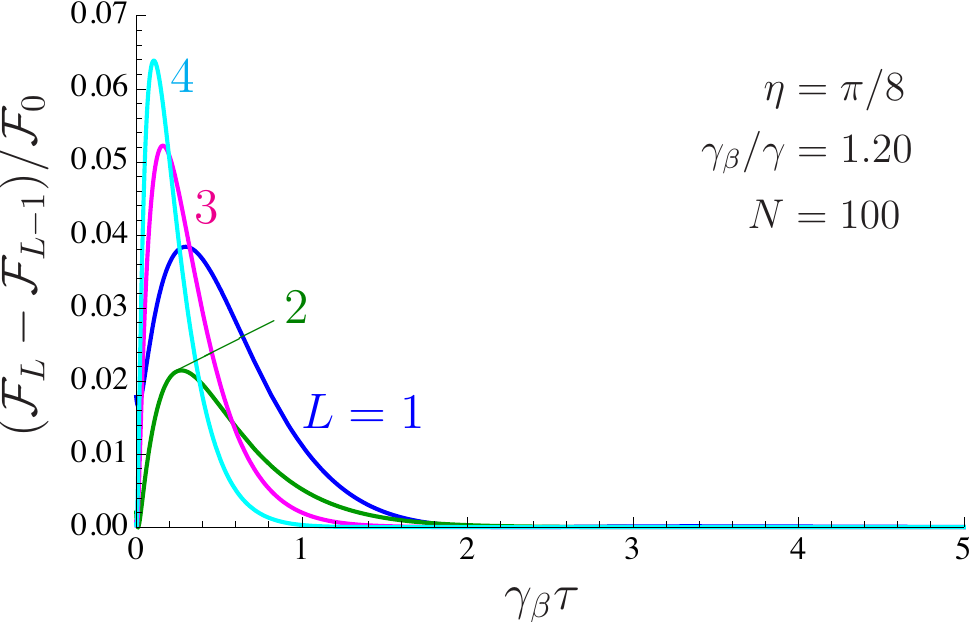}\\
\includegraphics[scale=0.53]{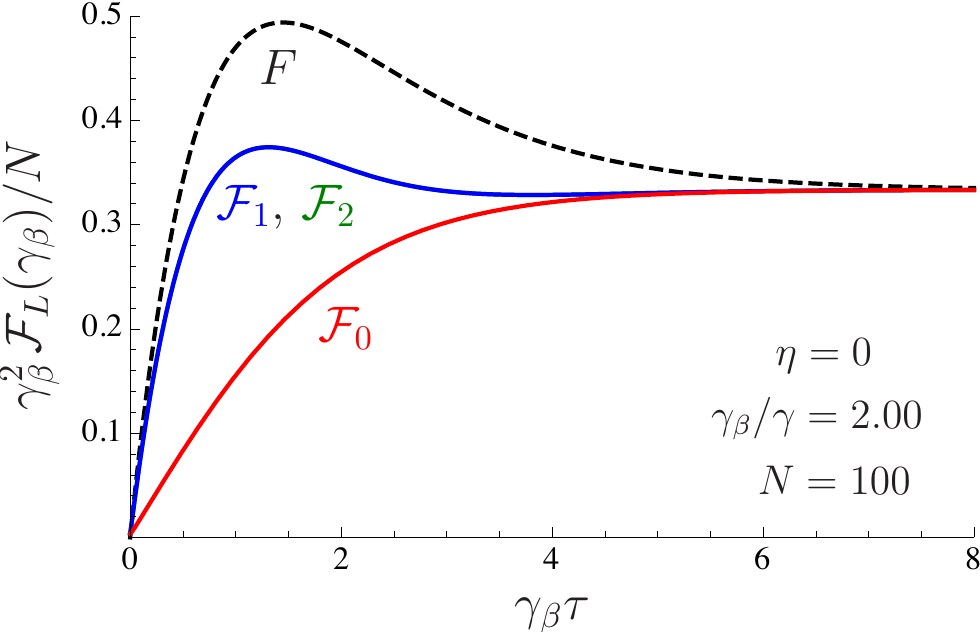}&
\includegraphics[scale=0.53]{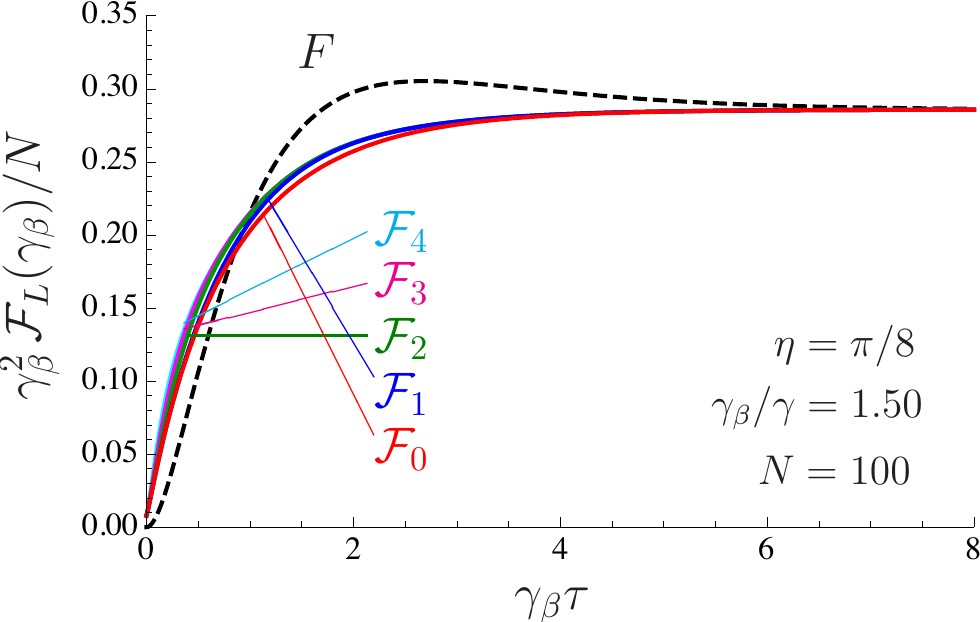}&
\includegraphics[scale=0.53]{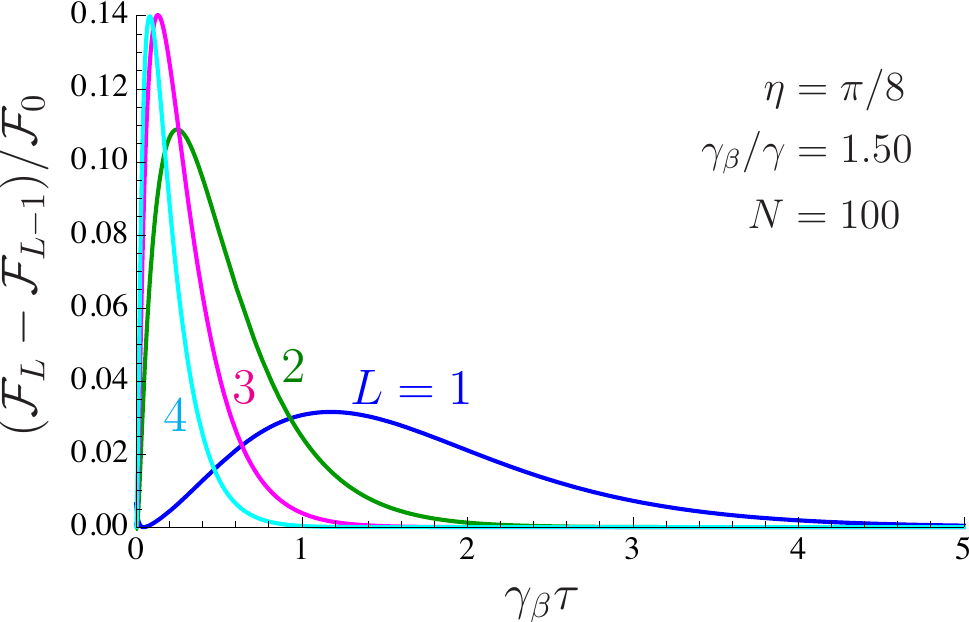}\\
\includegraphics[scale=0.53]{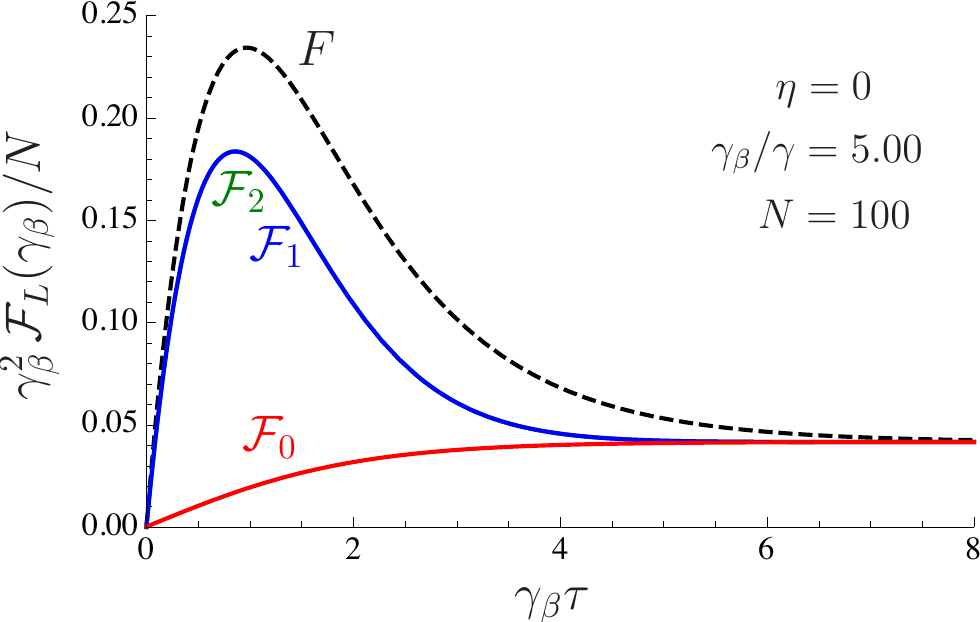}&
\includegraphics[scale=0.53]{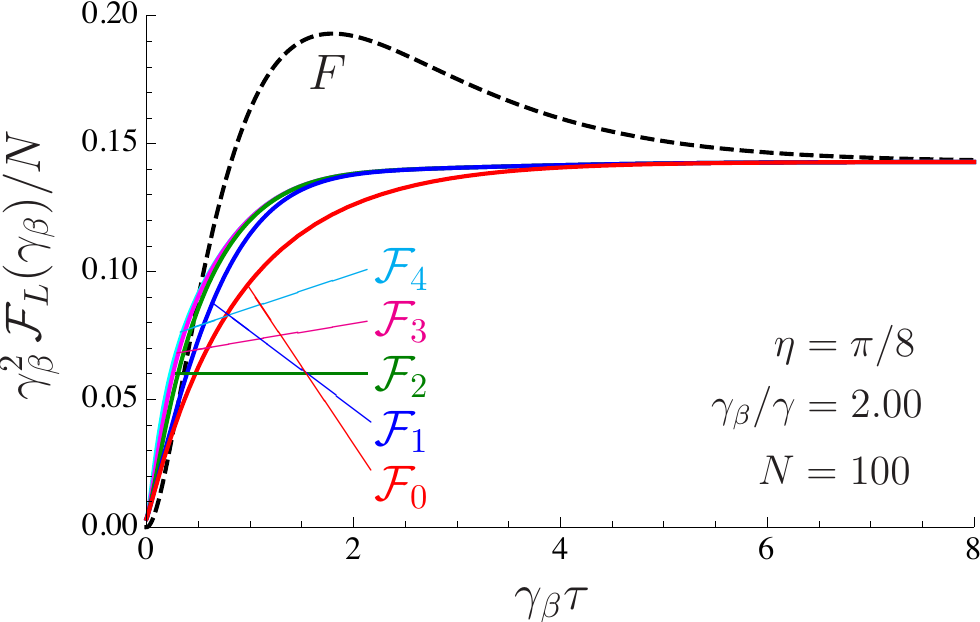}&
\includegraphics[scale=0.53]{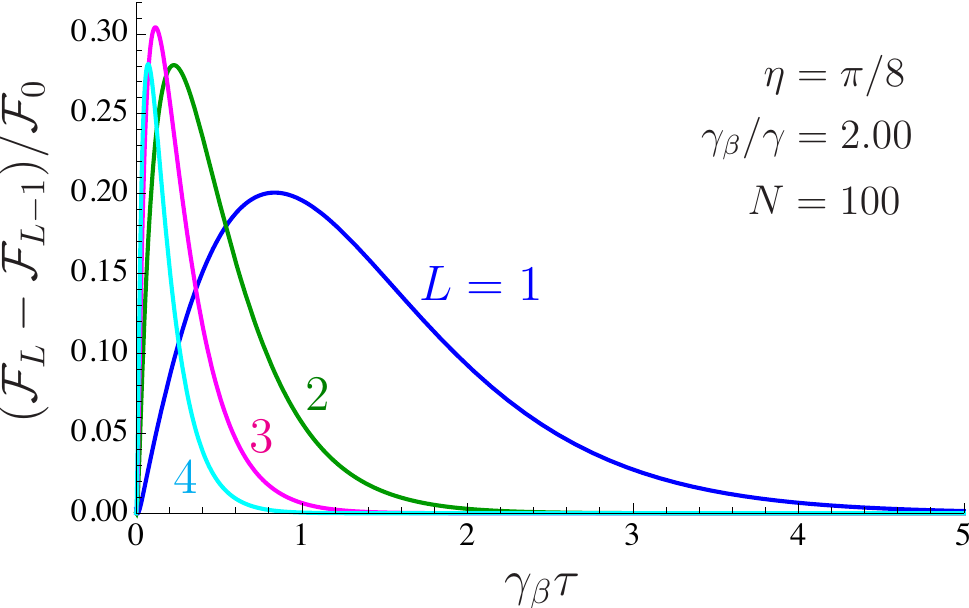}\\
\multicolumn{1}{r}{(b)\quad\ \mbox{}}&&\\[-3.5truemm]
\includegraphics[scale=0.53]{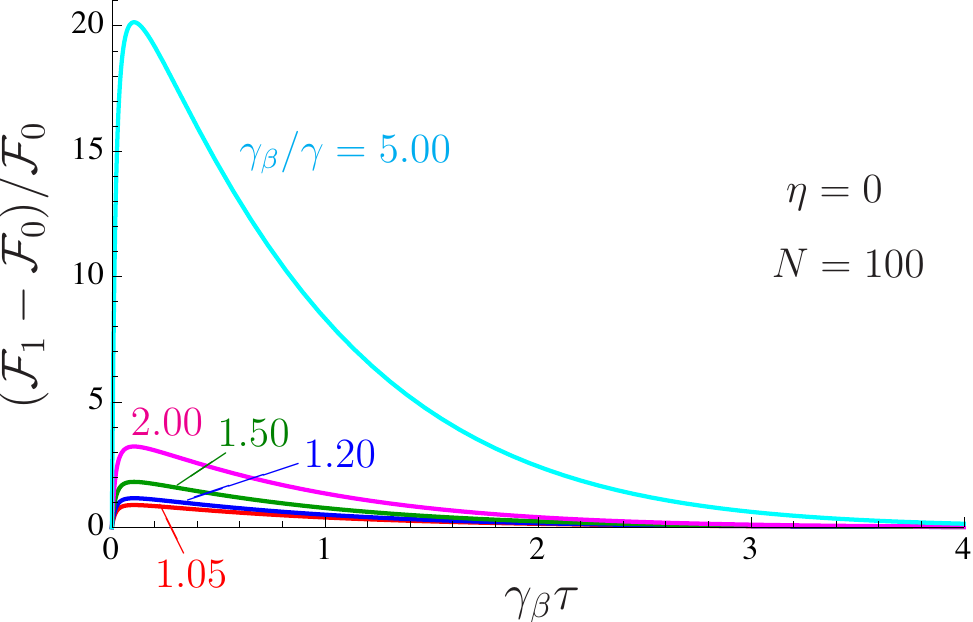}&
\includegraphics[scale=0.53]{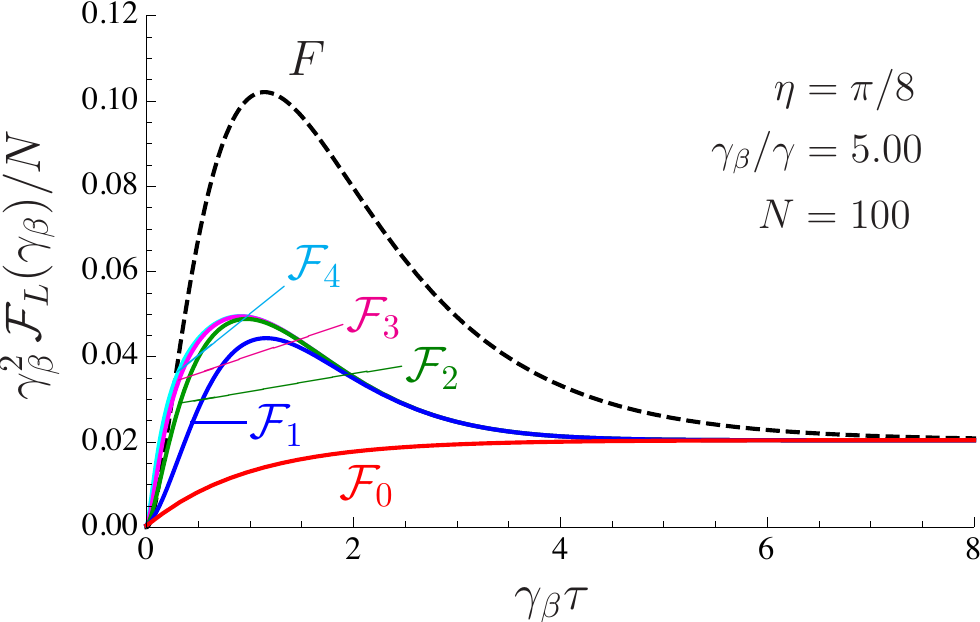}&
\includegraphics[scale=0.53]{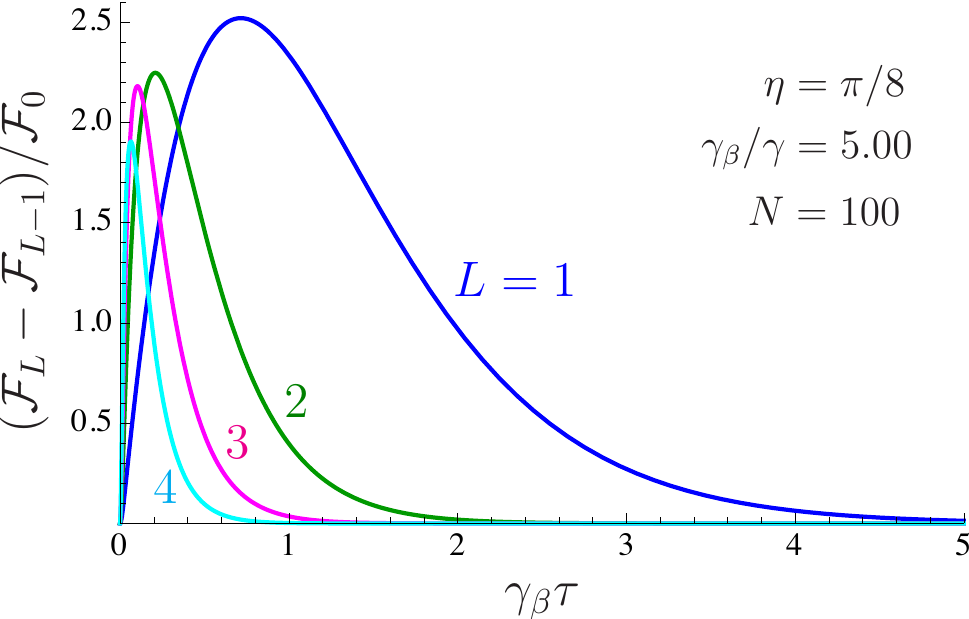}
\end{tabular}
\caption{The Fisher informations $\mathcal{F}_L(\gamma_\beta)/N$ per measurement by the sequential scheme (solid lines) are compared with the Fisher information $F(\gamma_\beta)$ by the standard strategy with $\rho_0=\ket{\downarrow}\bra{\downarrow}$ (dashed lines) for (a) projective measurements $\eta=0$ and for (c) weak measurements $\eta>0$.
The gains $[\mathcal{F}_L(\gamma_\beta)-\mathcal{F}_{L-1}(\gamma_\beta)]/\mathcal{F}_0(\gamma_\beta)$ by incorporating the correlation $C_L$ are shown in (b) for projective measurements $\eta=0$ and in (d) for weak measurements $\eta>0$.}
\label{fig:Fcomp}
\end{figure*}

\subsubsection{Projective Measurement $\eta=0$}
The Fisher informations $\mathcal{F}_L(\gamma_\beta)$ ($L=0,1,2$) by the sequential scheme are plotted in the five panels in Fig.\ \ref{fig:Fcomp}(a) and are compared with the Fisher information $F(\gamma_\beta)$ by the standard strategy with $\rho_0=\ket{\downarrow}\bra{\downarrow}$, for the case of projective measurements $\eta=0$.
In this case, the Fisher information $F(\gamma_\beta)$ coincides with the quantum Fisher information $F_Q(\gamma_\beta)$ in the standard strategy.

Compare first $F(\gamma_\beta)$ and $\mathcal{F}_0(\gamma_\beta)/N$ (per measurement).
We observe that the standard strategy provides better estimation than the sequential scheme.
Recall here that the input state $\rho_0=\ket{\downarrow}\bra{\downarrow}$ for $F(\gamma_\beta)$ is the optimal for the standard strategy. 
On the other hand, in the sequential scheme, the state of the qubit is projected into $\ket{\uparrow}$ or $\ket{\downarrow}$ depending on the outcome of the projective measurement.
If it is projected into $\ket{\uparrow}$ by a measurement, it restarts to evolve from this non-optimal state for the next measurement.
Not all the steps in the sequential measurements are optimal for the estimation.
That is why the sequential scheme cannot beat the standard strategy, in the case of projective measurement.

One can improve the performance of the sequential scheme, by incorporating $C_1$ for the estimation.
Indeed, as is clear from Fig.\ \ref{fig:Fcomp}(a), the Fisher information $\mathcal{F}_1(\gamma_\beta)$ for the estimation through $(S,C_1)$ is greater than the Fisher information $\mathcal{F}_0(\gamma_\beta)$ solely through $S$.
Note that no additional resources or experiments are required to incorporate $C_1$: one simply needs to carry out additional data analysis to compute $C_1$ from the data used to evaluate $S$.
In Fig.\ \ref{fig:Fcomp}(b), the gain in the Fisher information by incorporating $C_1$ is shown for different temperatures.

On the other hand, incorporating more correlation data, i.e., $C_\ell$ with $\ell>1$, does not help improve the estimation.
See Fig.\ \ref{fig:Fcomp}(a) again.
This is because every time one performs measurement the system is reset to a pure state by the projective measurement: there is no correlation between the measurement results separated over two steps.
The system simply repeats the same dynamics, jumping between pure states $\ket{\uparrow}$ and $\ket{\downarrow}$, and the measurement after multiple steps gains no more information than that attainable by the measurement after a single step.

\subsubsection{Weak Measurement $\eta>0$}
Let us next look at the cases with weak measurements $\eta>0$.
As is clear from Fig.\ \ref{fig:Fcomp}(c), the sequential scheme can be better than the standard strategy.
In particular, at low temperatures, the Fisher informations $\mathcal{F}_L(\gamma_\beta)/N$ ($L\ge0$) by the sequential scheme exceed the Fisher information $F(\gamma_\beta)$ by the standard strategy.

The reason is the following.
In the standard strategy, the weak measurement is performed only once, and the system is reset to the specific initial state $\rho_0$ for the next measurement.
The single weak measurement can acquire less information than a projective measurement, but if it is repeated many times, as in the sequential scheme, the information is accumulated, and better information is gained in our hands.
At the same time, the system is gradually projected to one of the eigenstates of the measured observable by the repeated weak measurements \cite{ref:ProgProj-Haroche}.
In other words, the repetition of the weak measurements mimics a stronger measurement (closer to a projective measurement).
That is why the sequential scheme can be better than the standard strategy, in the case of weak measurement.

It is also clear from Fig.\ \ref{fig:Fcomp}(c) that the precision of the estimation is improved by incorporating the correlation data $C_\ell$.
The gain in the Fisher information $[\mathcal{F}_L(\gamma_\beta)-\mathcal{F}_{L-1}(\gamma_\beta)]/\mathcal{F}_0$ by adding a correlation $C_L$ to $(S,C_1,\ldots,C_{L-1})$ is shown in Fig.\ \ref{fig:Fcomp}(d).
The enhancement is reminiscent when the time interval between measurements $\tau$ is short, i.e., $\gamma_\beta\tau\lesssim1$.
Moreover, the gain exhibits a peak at a smaller $\tau$ for a larger $L$.
This is because the two points of each two-point correlation $C_\ell$, separated by $\ell$ steps, should be within the correlation time $\tau_c\sim2/\gamma_\beta$ [which is ruled by the second largest eigenvalues $e^{-(\gamma_\beta/2\pm i\Omega)\tau}$ of the mixing channel (\ref{eqn:spectrum_model})], in order for the correlation $C_\ell$ to bear useful information.

It appears that the sequential scheme can beat the standard strategy only at low temperatures (small $\gamma_\beta$), but it should be noted that the standard strategy in Fig.\ \ref{fig:Fcomp} assumes the optimal initial state $\rho_0=\ket{\downarrow}\bra{\downarrow}$, while in the sequential scheme the system is around the stationary state $\rho_*$ of the mixing channel $\mathcal{E}$, which is the thermal equilibrium state $\rho_\text{eq}$ [see (\ref{eqn:FixedPointThEq})].
It would be more appropriate to compare the Fisher informations $\mathcal{F}_L(\gamma_\beta)/N$ by the sequential scheme with the Fisher information $F(\gamma_\beta)$ by the standard strategy in the large $\tau$ limit (which gives the Fisher information with the initial thermal state $\rho_0=\rho_\text{eq}$).

\section{Conclusions}\label{sec:conc}
The estimation of a parameter encoded in a quantum probe, through a series of measurements performed sequentially on the probe, has been analyzed in a general non-i.i.d.\ setting. 
On the basis of a diagrammatic approach we have discussed the conditions under which the central limit theorem holds as the number of measurements increases, reproducing the previous results \cite{ref:Guta-PRA} and generalizing them to the case where the correlations among the measurement data are also taken into account in the estimation strategy. 
Our analysis explicitly shows that the latter strategy can yield a significant advantage over the standard procedure where only the average of the acquired data is considered.

At present however it is not clear whether this is the best strategy one can do: it is indeed possible that different data processing (including the evaluation of higher-order correlations commented at the end of Sec.\ \ref{sec:Correlation}) can improve further the attainable accuracy. 
In the example studied in Sec.\ \ref{sec:ex}, the sequential scheme surpassed the standard i.i.d.\ procedure when we are able to perform only weak measurements, but could not beat the standard procedure when we are allowed to perform strong measurements.
A better strategy for the sequential scheme could beat the ultimate precision achievable by the standard strategy.
The optimal strategy would require different measurements step by step, or moreover would require quantum-correlated measurements over different measurement probings.
The use of entanglement is also an interesting possibility \cite{ref:RafalLorenzo-EntAgainstNoise}.
It is yet to be clarified what is the ultimate accuracy attainable in the sequential scheme for parameter estimation \cite{Note1}.

Recently, quantum metrology in the presence of noise is under intense study \cite{ref:NoiseMetroDavidovich-NatPhys,ref:Rafal-NatCommun,ref:RafalLorenzo-EntAgainstNoise,ref:Rafal-NJP2015}.
The mixing property required for the sequential scheme is relevant to noisy channels, and connections with such issue would be interesting to be explored.

\acknowledgments
This work is partially supported by the Top Global University Project from the Ministry of Education, Culture, Sports, Science and Technology (MEXT), Japan.
DB acknowledges support from the EPSRC Grant No.\ EP/M01634X/1.
KY is supported by a Grant-in-Aid for Scientific Research (C) (No.\ 26400406) from the Japan Society for the Promotion of Science (JSPS), and by a Waseda University Grant for Special Research Projects (No.\ 2015K-202).

\appendix
\section{Complete Expressions for the Covariances}
\label{app:CompCov}
In Sec.\ \ref{sec:Correlation}, we derived the asymptotic expression (\ref{eqn:MomentsAsymp}) for the even moments among $S$ and $C_\ell$. Here we provide the complete expressions for the covariances among $S$ and $C_\ell$ valid for any (even small) $N$.
Under the assumption that the quantum channel $\mathcal{E}$ is ergodic (not necessarily mixing), they read
\begin{widetext}
\begin{align}
\langle(S-\langle S\rangle_*)^2\rangle_N
={}&
\frac{1}{N}
(\Delta s)_*^2
+\frac{2}{N}
(1|
\tilde{\mathcal{E}}^{(1)}\frac{Q_*}{1-\mathcal{E}'}\tilde{\mathcal{E}}^{(1)}
|\rho_*)
-\frac{2}{N^2}
(1|
\tilde{\mathcal{E}}^{(1)}\frac{1-\mathcal{E}'^N}{(1-\mathcal{E}')^2}\mathcal{Q}_*\tilde{\mathcal{E}}^{(1)}
|\rho_*)
\nonumber\displaybreak[0]\\
&
{}-\frac{2}{N^2}
\sum_{j=1}^{N-1}
(1|
\tilde{\mathcal{E}}^{(1)}\frac{\mathcal{E}'^{N-j}}{1-\mathcal{E}'}\mathcal{Q}_*\tilde{\mathcal{E}}^{(1)}\mathcal{E}'^{j-1}\mathcal{Q}_*
|\rho_0)
+\frac{1}{N^2}
(1|
\tilde{\mathcal{E}}^{(2)}\frac{1-\mathcal{E}'^N}{1-\mathcal{E}'}\mathcal{Q}_*
|\rho_0)
\nonumber\displaybreak[0]\\
&
{}+\frac{2}{N^2}
(1|
\tilde{\mathcal{E}}^{(1)}\frac{\mathcal{Q}_*}{1-\mathcal{E}'}\tilde{\mathcal{E}}^{(1)}\frac{1-\mathcal{E}'^{N-1}}{1-\mathcal{E}'}\mathcal{Q}_*
|\rho_0)
\qquad\qquad\qquad\qquad\qquad
(N\ge1),
\label{eqn:DeltaS2Complete}
\end{align}
\begin{align}
&\langle(C_\ell-\langle C_\ell\rangle_*)(C_{\ell'}-\langle C_{\ell'}\rangle_*)\rangle_N
\nonumber\\
&\quad
=
\frac{\theta(N-\ell-\ell')}{(N-\ell)(N-\ell')}
\,\Biggl[
(1|
\tilde{\mathcal{E}}_{\ell}^{(1)}
\left(
\frac{N-\ell-\ell'}{1-\mathcal{E}'}
-\frac{1-\mathcal{E}'^{N-\ell-\ell'}}{(1-\mathcal{E}')^2}
\right)
\mathcal{Q}_*
\tilde{\mathcal{E}}_{\ell'}^{(1)}
|\rho_*)
+(\ell\leftrightarrow\ell')
+
(N-\ell-\ell')
(1|
\tilde{\mathcal{E}}_{\ell\ell'}^{\circ\star\bullet}
|\rho_*)
\Biggr]
\nonumber\displaybreak[0]\\
&\qquad
{}+
\frac{1}{(N-\ell)(N-\ell')}
\sum_{k=\max(1+\ell+\ell'-N,1)}^{\min(\ell,\ell')-1}
(N-\ell-\ell'+k)
(1|
\tilde{\mathcal{E}}_{\ell\ell',k}^{\circ\bullet\circ\bullet}
|\rho_*)
\nonumber\displaybreak[0]\\
&\qquad
{}+
\frac{1}{N-\min(\ell,\ell')}
(1|
\delta_{\ell\ell'}
\tilde{\mathcal{E}}_{\ell}^{\star\star}
+(1-\delta_{\ell\ell'})
(
\tilde{\mathcal{E}}_{\ell\ell'}^{\star\circ\bullet}
+
\tilde{\mathcal{E}}_{\ell\ell'}^{\bullet\circ\star}
+
\tilde{\mathcal{E}}_{\ell\ell'}^{\bullet\circ\circ\bullet}
)
|\rho_*)
\quad
[\rho_0=\rho_*,\ \ell,\ell'\ge1,\ N\ge\max(\ell,\ell')+1],
\end{align}
and
\begin{align}
\langle(S-\langle S\rangle_*)(C_\ell-\langle C_\ell\rangle_*)\rangle_N
=
\frac{1}{N}
(1|
\tilde{\mathcal{E}}_\ell^{(2)}
|\rho_*)
&
{}+\frac{1}{N(N-\ell)}
(1|\tilde{\mathcal{E}}^{(1)}
\left(
\frac{N-\ell}{1-\mathcal{E}'}
-\frac{1-\mathcal{E}'^{N-\ell}}{(1-\mathcal{E}')^2}
\right)
\mathcal{Q}_*
\tilde{\mathcal{E}}_\ell^{(1)}
|\rho_*)
\nonumber\\
&
{}+\frac{1}{N(N-\ell)}
(1|\tilde{\mathcal{E}}_\ell^{(1)}
\left(
\frac{N-\ell}{1-\mathcal{E}'}
-\frac{1-\mathcal{E}'^{N-\ell}}{(1-\mathcal{E}')^2}
\right)
\mathcal{Q}_*
\tilde{\mathcal{E}}^{(1)}
|\rho_*)
\nonumber\\
&
\qquad\qquad\qquad\qquad\qquad\qquad\qquad
(\rho_0=\rho_*,\ \ell\ge1,\ N\ge\ell+1),
\end{align}
where $\tilde{\mathcal{E}}^{(m)}$ and $(\Delta s)_*^2$ are defined in (\ref{eqn:En}) and (\ref{eqn:DeltasStar}), respectively, $\tilde{\mathcal{E}}_\ell^{(m)}$ are in (\ref{eqn:E1l}) and (\ref{eqn:E2l}), and the other components are given in (\ref{eqn:E2llcomp1})--(\ref{eqn:E2llcomp6}).

\section{Covariances among $\bm{C_\ell}$ for the Model}
\label{sec:covariances}
In (\ref{eqn:CovCModelAsymp}) in Sec.\ \ref{sec:ex} we showed the asymptotic expression for the covariance between $C_\ell$ and $C_{\ell'}$ for large $N$ for the model. Here we provide its complete expression valid for any (even small) $N$.
In the stationary state $\rho_0=\rho_*$, the covariances between $C_\ell$ and $C_{\ell'}$ ($\ell\ge\ell'\ge1$) are given for $N\ge\ell+\ell'$ by
\begin{align}
&\langle C_\ell C_{\ell'}\rangle_N-\langle C_\ell\rangle_N\langle C_{\ell'}\rangle_N
\nonumber\\
&\quad
=\delta_{\ell\ell'}
\frac{1}{N-\ell'}
\sin^4\!2\eta
+
\frac{2}{N-\ell'}
\left[
\langle C_{\ell-\ell'}\rangle_*
+\left(
1-\frac{\ell'}{N-\ell}
\right)
\langle C_{\ell+\ell'}\rangle_*
\right]
\sin^2\!2\eta
\nonumber\displaybreak[0]\\
&\qquad
{}-
\frac{1}{N-\ell'}
\,\Biggl\{
\ell'\left(
2-\frac{\ell'}{N-\ell}
\right)
e^{-(\ell+\ell')\gamma_\beta\tau}
-
\frac{
1+e^{-2 \gamma_\beta\tau}
}{
1-e^{-2 \gamma_\beta\tau}
}
\left[
e^{-(\ell-\ell')\gamma_\beta\tau}
-\left(
1-\frac{\ell'}{N-\ell}
\right)
e^{-(\ell+\ell')\gamma_\beta\tau}
\right]
\nonumber\displaybreak[0]\\
&\qquad\qquad\qquad\qquad\qquad\qquad\qquad
{}-\left(
\ell-\ell'
-\frac{2}{N-\ell}
\frac{
e^{-2 \gamma_\beta\tau }
}{
(1-e^{-2 \gamma_\beta\tau })^2
}
\right)
(
e^{-(\ell-\ell')\gamma_\beta\tau}
-e^{-(\ell+\ell')\gamma_\beta\tau}
)
\Biggr\}
\left(
1-\frac{\gamma^2}{\gamma_\beta^2}
\right)^2
\cos^4\!2\eta
\nonumber\displaybreak[0]\\
&\qquad
{}-\frac{2}{N-\ell'}
\,\Biggl\{
\ell'\left(
2-\frac{\ell'}{N-\ell}
\right)
(
e^{-\ell\gamma_\beta\tau}
+e^{-\ell'\gamma_\beta\tau}
)
-\frac{
1+e^{-\gamma_\beta\tau}
}{
1-e^{-\gamma_\beta\tau}
}
\left[
\left(
2-\frac{\ell'}{N-\ell}
\right)
(
1
-e^{-\ell'\gamma_\beta\tau}
)
+\frac{\ell'}{N-\ell}
e^{-\ell\gamma_\beta\tau}
\right]
\nonumber\displaybreak[0]\\
&\qquad\qquad\qquad\quad
{}
+\frac{1}{N-\ell}
\frac{
e^{-\gamma_\beta\tau}
}{(1-e^{-\gamma_\beta\tau})^2}
\,\Bigl[
(1-e^{-(N-\ell-\ell')\gamma_\beta\tau})
(1-e^{-\ell\gamma_\beta\tau})
(1-e^{-\ell'\gamma_\beta\tau})
+e^{-(\ell-\ell')\gamma_\beta\tau}
-e^{-(\ell+\ell')\gamma_\beta\tau}
\Bigr]
\Biggr\}\,
\nonumber\\
&\qquad\qquad\qquad\qquad\qquad\qquad\qquad\qquad\qquad\qquad\qquad\qquad\qquad\qquad\qquad
\qquad\qquad\qquad\qquad\quad\ %
{}\times
\left(
1-\frac{\gamma^2}{\gamma_\beta^2}
\right)
\frac{\gamma^2}{\gamma_\beta^2}
\cos^4\!2\eta,
\end{align}
while for $\ell+\ell'>N\ge\ell+1$ by
\begin{align}
&\langle C_\ell C_{\ell'}\rangle_N-\langle C_\ell\rangle_N\langle C_{\ell'}\rangle_N
\nonumber\\
&\ \ %
=\delta_{\ell\ell'}
\frac{1}{N-\ell'}
\sin^4\!2\eta
+\frac{2}{N-\ell'}
\langle C_{\ell-\ell'}\rangle_*
\sin^2\!2\eta
\nonumber\\
&\quad\ \ %
{}-
\left[
e^{-(\ell+\ell')\gamma_\beta\tau}
-\frac{1}{N-\ell'}
\left(
\ell-\ell'
+\frac{
1+e^{-2 \gamma_\beta\tau}
}{
1-e^{-2 \gamma_\beta\tau}
}
-
\frac{2}{N-\ell}
\frac{
1-e^{-2(N-\ell)\gamma_\beta\tau}
}{
(1-e^{-2 \gamma_\beta\tau })^2}
e^{-2\gamma_\beta\tau}
\right)
e^{-(\ell-\ell')\gamma_\beta\tau}
\right]
\left(
1-\frac{\gamma^2}{\gamma_\beta^2}
\right)^2
\cos^4\!2\eta
\nonumber\\
&\quad\ \ %
{}-2\,\Biggl[
\left(
1-\frac{\ell-\ell'}{N-\ell'}
\right)
(e^{-\ell\gamma_\beta\tau}+e^{-\ell'\gamma_\beta\tau}) 
-\frac{1}{N-\ell'}
\frac{
1+e^{-\gamma_\beta\tau}
}{
1-e^{-\gamma_\beta\tau}
}
(1
+
e^{-\ell\gamma_\beta\tau}-e^{-\ell'\gamma_\beta\tau}
) 
\nonumber\\
&\quad\qquad\quad\ \ %
{}+\frac{1}{(N-\ell)(N-\ell')}
\frac{
1
-e^{-(N-\ell)\gamma_\beta\tau}
}{
(1-e^{-\gamma_\beta\tau})^2
}
e^{-\gamma_\beta\tau}
(1+e^{-(\ell-\ell')\gamma_\beta\tau}
-
e^{(N-\ell-\ell')\gamma_\beta\tau}
+
e^{-\ell\gamma_\beta\tau}
)
\Biggr]
\left(
1-\frac{\gamma^2}{\gamma_\beta^2}
\right)
\frac{\gamma^2}{\gamma_\beta^2}
\cos^4\!2\eta.
\end{align}
\end{widetext}


\begin{thebibliography}{10}

\bibitem{BENSHOR}
C. H. Bennett and P. W. Shor, IEEE Trans. Inf. Theory \textbf{44},  2724  (1998).

\bibitem{ref:QuantumMetrologyVittorio}
V. Giovannetti, S. Lloyd, and L. Maccone, Phys. Rev. Lett. \textbf{96},  010401
   (2006).

\bibitem{ref:MetrologyNaturePhoto}
V. Giovannetti, S. Lloyd, and L. Maccone, Nat. Photon. \textbf{5},  222
  (2011).

\bibitem{VIRGO}
L.~P. Singer, L.~R. Price, B. Farr, A.~L. Urban, C. Pankow, S. Vitale, J.
  Veitch, W.~M. Farr, C. Hanna, K. Cannon, T. Downes, P. Graff, C.-J. Haster,
  I. Mandel, T. Sidery, and A. Vecchio, Astrophys. J. \textbf{795},  105
  (2014).

\bibitem{AUZ}
M. Auzinsh, D. Budker, D.~F. Kimball, S.~M. Rochester, J.~E. Stalnaker, A.~O.
  Sushkov, and V.~V. Yashchuk, Phys. Rev. Lett. \textbf{93},  173002  (2004).

\bibitem{MAG1}
F. Wolfgramm, C. Vitelli, F.~A. Beduini, N. Godbout, and M.~W. Mitchell, Nat.
  Photon. \textbf{7},  28  (2013).

\bibitem{ref:Helstrom}
C.~W. Helstrom, \textit{Quantum Detection and Estimation Theory} (Academic
  Press, New York, 1976).

\bibitem{ref:BraunsteinCave1994}
S.~L. Braunstein and C.~M. Caves, Phys. Rev. Lett. \textbf{72},  3439  (1994).

\bibitem{ref:BraunsteinCave1996AnnPhys}
S.~L. Braunstein, C.~M. Caves, and G. J. Milburn, Ann. Phys. (N.Y.) \textbf{247},
  135  (1996).

\bibitem{ref:QuantumEstimation}
\textit{Quantum State Estimation}, edited by M.~G.~A. Paris and J.
  \v{R}eh\'{a}\v{c}ek (Springer, Berlin, 2004).

\bibitem{ref:Paris-IJQI}
M.~G.~A. Paris, Int. J. Quant. Inf. \textbf{7},  125  (2009).

\bibitem{ref:HolevoSNS}
A.~S. Holevo, \textit{Probabilistic and Statistical Aspects of Quantum Theory}
  (Edizioni della Normale, Pisa, 2011).

\bibitem{ref:TerhalDiVincenzo}
B.~M. Terhal and D.~P. DiVincenzo, Phys. Rev. A \textbf{61},  022301  (2000).

\bibitem{ref:Mixing-Wolf}
M.~M. Wolf, ``Quantum Channels \& Operations: Guided Tour,'' URL:
  https://www-m5.ma.tum.de/foswiki/pub/
  M5/Allgemeines/MichaelWolf/QChannelLecture.pdf.

\bibitem{ref:MixingNJP-BurgarthGiovannetti}
D. Burgarth and V. Giovannetti, New J. Phys. \textbf{9},  150  (2007).

\bibitem{ref:ConvexErgodicity}
D. Burgarth, G. Chiribella, V. Giovannetti, P. Perinotti, and K. Yuasa, New J.
  Phys. \textbf{15},  073045  (2013).

\bibitem{ref:Mabuchi-QSO1996}
H. Mabuchi, Quant. Semiclass. Opt. \textbf{8},  1103  (1996).

\bibitem{ref:GambettaWiseman-PRA2001}
J. Gambetta and H.~M. Wiseman, Phys. Rev. A \textbf{64},  042105  (2001).

\bibitem{ref:Guta-PRA}
M. Gu\c{t}\u{a}, Phys. Rev. A \textbf{83},  062324  (2011).

\bibitem{ref:Guta-ContTime}
C. C\u{a}tan\u{a}, M. van Horssen, and M. Gu\c{t}\u{a}, Phil. Trans. R. Soc. A
  \textbf{370},  5308  (2012).

\bibitem{ref:Molmer-PRA2013}
S. Gammelmark and K. M\o{}lmer, Phys. Rev. A \textbf{87},  032115  (2013).

\bibitem{ref:Molmer-PRL2014}
S. Gammelmark and K. M\o{}lmer, Phys. Rev. Lett. \textbf{112},  170401  (2014).

\bibitem{ref:RybarZiman}
T. Ryb\'{a}r and M. Ziman, Phys. Rev. A \textbf{92}, 042315 (2015).

\bibitem{ref:GutaHorssen-JMP2015}
M. van Horssen and M. Gu\c{t}\u{a}, J. Math. Phys. \textbf{56}, 022109 (2015).

\bibitem{ref:ZenoEst}
A.~H. Kiilerich and K. M\o{}lmer, Phys. Rev. A \textbf{92},  032124  (2015).

\bibitem{HOPFNER}
R. H\"opfner, J. Jacod, and L. Ladelli, Probab. Th. Rel. Fields \textbf{86},
  105  (1990).

\bibitem{ref:DasGuptaCh10}
A. DasGupta, \textit{Asymptotic Theory of Statistics and Probability}
  (Springer, New York, 2008), Chap.~10.

\bibitem{ref:BillingsleyTh27.4}
P. Billingsley, \textit{Probability and Measure}, {Anniversary} ed. (Wiley, New
  York, 2012), {Theorem 27.4}.

\bibitem{ref:NielsenChuang}
M.~A. Nielsen and I.~L. Chuang, \textit{Quantum Computation and Quantum
  Information} (Cambridge University Press, Cambridge, 2000).

\bibitem{ref:DynamicalMap-Alicki}
R. Alicki and K. Lendi, \textit{Quantum Dynamical Semigroups and Applications},
  2nd  ed. (Springer, Berlin, 2007).

\bibitem{ref:HolevoVittorioReview}
A.~S. Holevo and V. Giovannetti, Rep. Prog. Phys. \textbf{75},  046001  (2012).

\bibitem{ref:GeometryOfQuantumStates}
I. Bengtsson and K. {\.Z}yczkowski, \textit{Geometry of Quantum States: An
  Introduction to Quantum Entanglement} (Cambridge University Press, Cambridge,
  2006).

\bibitem{ref:Cramer}
H. Cram\'er, \textit{Mathematical Methods of Statistics} (Princeton University
  Press, Princeton, 1946).

\bibitem{GERARDO}
L.~A. Correa, M. Mehboudi, G. Adesso, and A. Sanpera, Phys. Rev. Lett.
  \textbf{114},  220405  (2015).

\bibitem{ref:LocalThermometerAnto}
A. {De Pasquale}, D. Rossini, R. Fazio, and V. Giovannetti, arXiv:1504.07787
  [quant-ph]  (2015).

\bibitem{ref:QuantumOptics-Scully}
M.~O. Scully and M.~S. Zubairy, \textit{Quantum Optics} (Cambridge University
  Press, Cambridge, 1997).

\bibitem{ref:OpenQuantumSystems}
H.-P. Breuer and F. Petruccione, \textit{The Theory of Open Quantum Systems}
  (Oxford University Press, Oxford, 2002).

\bibitem{ref:QuantumNoise}
C. W. Gardiner and P. Zoller, \textit{Quantum Noise}, 3rd  ed. (Springer, Berlin,
  2004).

\bibitem{ref:QuantumOptics-WallsMilburn}
D.~F. Walls and G.~J. Milburn, \textit{Quantum Optics}, 2nd  ed. (Springer,
  Berlin, 2008).

\bibitem{ref:Wiseman-NJP-Uncertainty}
A.~P. Lund and H.~M. Wiseman, New J. Phys. \textbf{12},  093011  (2010).

\bibitem{ref:Steinberg-PRL-Uncertainty}
L.~A. Rozema, A. Darabi, D.~H. Mahler, A. Hayat, Y. Soudagar, and A.~M.
  Steinberg, Phys. Rev. Lett. \textbf{109},  100404  (2012).

\bibitem{ref:QFI-convexity-Fujiwara2001}
A. Fujiwara, Phys. Rev. A \textbf{63},  042304  (2001).

\bibitem{ref:ProgProj-Haroche}
C. Guerlin, J. Bernu, S. Del\'{e}glise, C. Sayrin, S. Gleyzes, S. Kuhr, M.
  Brune, J.-M. Raimond, and S. Haroche, Nature (London) \textbf{448},  889
  (2007).

\bibitem{ref:RafalLorenzo-EntAgainstNoise}
R. Demkowicz-Dobrza\ifmmode~\acute{n}\else \'{n}\fi{}ski and L. Maccone, Phys.
  Rev. Lett. \textbf{113},  250801  (2014).

\bibitem{Note1}
In Ref.\ \cite {ref:Guta-PRA} a concise formula for the quantum Fisher
  information providing the maximum Fisher information attainable by the
  optimal measurement strategy applied on the target system plus probing
  systems in the sequential scheme is derived. It is however valid only when
  each probing process is unitary. Moreover, it is not applicable if one can
  measure the target system only through the probes.

\bibitem{ref:NoiseMetroDavidovich-NatPhys}
B.~M. Escher, R.~L. de~Matos~Filho, and L. Davidovich, Nat. Phys. \textbf{7},
  406  (2011).

\bibitem{ref:Rafal-NatCommun}
R. Demkowicz-Dobrza\ifmmode~\acute{n}\else \'{n}\fi{}ski, J. Ko\l ody\ifmmode~\acute{n}\else \'{n}\fi{}ski, and M.
  Gu\c{t}\u{a}, Nat. Commun. \textbf{3},  1063  (2012).

\bibitem{ref:Rafal-NJP2015}
M. Jarzyna and R. Demkowicz-Dobrza\ifmmode~\acute{n}\else \'{n}\fi{}ski, New J.
  Phys. \textbf{17},  013010  (2015).

\end{thebibliography}
\end{document}